\documentclass{ws-ijmpe}
\usepackage[super,compress]{cite}
\usepackage{graphicx}
\usepackage{multirow}

\graphicspath{ {./images/} }
\usepackage{subcaption}
\usepackage{hyperref}
\hypersetup{
linkcolor=black,
citecolor=black
}
\begin{document}

\markboth{E. Shokr, A. H. El-Farrash, A. De Roeck, and M.A. Mahmoud  }{Study of charged-particle multiplicity fluctuations  in $pp$ collisions}

%\catchline{}{}{}{}{}
%%%%%%%%%%%%%%%%%%%%%%%%%%%%%%%%%%%%%%%%%%%%%%%%%%%%%%%%%%%%%%%%%%%%

\title{Study of charged-particle multiplicity fluctuations  in $pp$ collisions with Monte Carlo event generators at the LHC}

%\author{E. Shokr*}
\author{{E. Shokr\footnote{Corresponding author. \newline
preprint submitted to the International Journal of Modern Physics E
}}   $ $ and A. H. El-Farrash}

\address{Physics Department, Faculty of Science, Mansoura University, Egypt.\\
eslam.shokr@cern.ch}

%\author{A. H. El-Farrash}

%\address{Physics Department, Faculty of Science, Mansoura University, Egypt.\\
%ali\_elfarrash@hotmail.com
%}

\author{A. De Roeck}

\address{CERN, Geneva, Switzerland.\\
%albert.de.Roeck@cern.ch
}

\author{M.A. Mahmoud}

\address{Center for High Energy Physics (CHEP-FU), Faculty of Science, Fayoum University, Egypt.\\
%mohammed.attia@cern.ch
}

\maketitle
\begin{history}
\received{Day Month Year}
\revised{Day Month Year}
%\accepted{Day Month Year}
%\comby{(xxxxxxxxxx)}
\end{history}

\begin{abstract}
Proton-Proton ($pp$) collisions at the Large Hadron Collider (LHC) are simulated  in order to study events with a high local density of charged particles produced in narrow pseudorapidty windows  of $\Delta\eta$ = 0.1, 0.2, and 0.5. The $pp$ collisions are generated at center of mass energies of $\sqrt{s} = 2.36$, $7$, $8$, and $13$ TeV, i.e. the energies at which the LHC has operated so far, using PYTHIA and HERWIG event generators.
We have also studied  the average of the maximum charged-particle density versus the event multiplicity for all events, using the different pseudorapidity windows.
This study prepares for the multi-particle production background expected in a future search for anomalous
high-density 
multiplicity fluctuations using the LHC data.
\end{abstract}


%\ccode{PACS numbers:}

%\tableofcontents

\section{Introduction}
\label{sec1}

The advent of the Large Hadron Collider (LHC) on the 
scientific 
scene offers a plethora of  opportunities for  
searches for new physics phenomena 
beyond those produced in the current
models and theories. 
While searching for physics beyond the Standard
Model have been an intense study area using the LHC data, 
during the last 10 years, for hunting  new particles or 
interactions, here we are interested in 
the 
potential to search for the production of 
anomalous high-density “spikes" of the number of particles 
produced in narrow pseudorapidity intervals.
%Such events, with a large concentration of charged particles in small (pseudo)rapidity windows, have been searched for in a 
%number of experiments.
%e.g. see\cite{%K.I.Aleksejeva,
%N. Arata, N.A. Marutyan, G.A. Apanasenko,T.H. Burnatt}, and
%consecutively has in hadronic collisions for a long time by %several experimental collaborations.
Events 
producing such anomalous particle clusters  are  called ring-like or spike events\cite{I.M. Dremin}, since the particles typically
tend to be emitted with nearly equal polar angle but extend in  the full azimuthal angular phase space region, hence
they can form a ring of particles in a plane perpendicular to the axis of collision.

First results on the observation  of such multiplicity fluctuations were reported by cosmic rays experiments, see e.g.~\cite{N. Arata, T.H. Burnatt,G.A. Apanasenko}, and
these somewhat unexpected results have led to the suggestion that such spikes of particles  could result from the formation of 
“hot-spots" of matter, particularly in 
heavy ion collisions\cite{Hwa1987} and
possibly in the form of a Quark-Gluon Plasma (QGP)\cite{GWLASSY},  i.e. a dense state of quarks and 
gluons, 
where these gluons and quarks are no longer confined inside well-separated hadrons. Such new state of matter has been 
observed meanwhile at heavy ion experiments at the Relativistic Heavy Ion Collider (RHIC) and 
the LHC.

The UA5 collaboration~\cite{P. Carlson}   used a sample of 
so-called minimum bias events in proton-antiproton ($p\overline{p}$) collisions to search for signs of very high particle density fluctuations as a potential signal for new physics~\cite{L. Van Hove}. 
They observed events with 15 charged particles or more in a pseudorapidity window ($\Delta\eta) = $ 0.5, in
$p\overline{p}$ 
collisions at center of mass (C.M.) energy of 540 GeV.
%, which did not have any noticeable jet structure.  
Here, $\Delta\eta$ is the difference in pseudorapidity ($\eta$) between two charged particles and $\eta = - \ln(\tan(\theta /2) $, with $\theta$  the polar angle with respect to the beam axis.

In $p\overline{p}$ and $pp$ collisions such formation of a new state of matter 
would however be highly unexpected, but
other possible ideas for the origin of such effect
in terms of coherent Quantum Chromodynamics (QCD) hadron production \cite{G.A. Apanasenko, Dremin1979,Dremin2005}, analogous to well known 
Quantum Electrodynamics (QED) Cherenkov radiation, or  Mach waves \cite{ Dremin2006} have been proposed as well.

Inspired by the UA5 observation, the NA22 collaboration~\cite{Adamus:1986vd} at the CERN's  Super Proton Synchrotron (SPS) accelerator conducted a search for such spike events in $K^+ p$, $\pi^+ p$, and $pp$ collisions at $\sqrt{s} = 22$ GeV. They studied  the 
distribution of events 
versus the  maximum charged-particle density 
($n_{ch}$) found in a rapidity window ($\Delta y$) = 0.1,  
and reported 
that the number of events ($N$) decreases with increasing of $n_{ch}$, following an 
approximate exponential 
dependence $dN/dn_{ch} = a$ $exp(-bn_{ch})$.
%with slope $b$ and constant $a$. 
In addition to the smooth exponential decrease of the data, they observed a single 
outlying anomalous event with 10 charged particles within a very narrow rapidity window of $\Delta y = 0.098$, corresponding to a density of 
about 100 particles per unit of  rapidity. This is
a higher
density than what was  observed by the UA5 experiment, despite that the 
C.M. energy for the collisions at the UA5 is 
about 25 times larger. The anomalous event is reported to have
an expectation probability of $10^{-3}$ which calculated from the exponential 
fit to the data, and the cluster of tracks is located very
centrally in the event, near rapidity (y) = 0.
The 10 particles are very uniformly distributed in the azimuthal angle ($\phi$), and hence they don't show the typical characteristics 
of a mini-jet formation.
%the collaborators refereed this event as an anomalous event.

%done_ smae paper parameters
J.B. Singh and J.M. Kohli  \cite{J.B. Singh} have studied the distribution of events with maximum charged ($n_{ch}$)
and total ($n_{tot}$) particle density located within $\Delta y$ = 0.1 in $pp$ collisions at $\sqrt{s} = 26$ GeV, based on the data recorded by 
the NA23 experiment.
They observed events with a high local particle density of about 100 particles per unit rapidity but only when  both charged and neutral particles are included. The exponential behavior was observed but no outlying anomalous events were reported, based
on the total event statistics of NA23 which is a factor of four smaller %compared to 
than the one of NA22.

The NA22 collaboration further studied the average of the maximum number of charged particles $<(\Delta n / \Delta\eta)_{max}>$, in $\Delta\eta = $ 0.1 and 0.5,  as a function  of the full event multiplicity,
i.e. the total number of charged particles produced in the  collision. 
%which is in this case a $\pi^+ p$ collision. 
The 
result for 
 $\Delta\eta = $ 0.5  at $\sqrt{s} = 22$ GeV for $\pi^+ p$ was compared with the UA5 data\cite{Geich_Gimbel} at $\sqrt{s} =$  540 GeV for the same pseudorapidity window. The $<(\Delta n / \Delta\eta)_{max}>$ distribution has an approximate linear dependence on multiplicity and was found to be almost independent on the energy between 200 GeV to 900 GeV. The NA22 data support this observation and extended the range from 900 GeV down to 22 GeV. Several more studies were done on this topic at different C.M. energies \cite{J.B. Singh, N.M. Agababyan, Fabbri:192279}, leading to similar findings.

As mentioned, one possible interpretation to explain such large density fluctuations and the production of events with anomalously
high-density, as has been proposed, is the formation of the so-called 
hot-spots in a new state of matter, would
not be that this is, really, only expected in nucleus-nucleus
collisions and not in more simple  $pp$, $p\overline{p}$, or meson-proton collisions.
Interestingly, in 2010, at the startup of the LHC at CERN, Geneva, Switzerland, when the first 
$pp$ collisions at C.M. energy of 7 TeV
were produced, the highest energy produced in the laboratory at that 
time, unexpected correlations were observed in minimum bias and high multiplicity events by the CMS experiment which were until then thought
to be only produced in nucleus-nucleus collisions
\cite{Khachatryan:2010gv} and until then had been linked  to the possible formation of
QGP. 
The same intriguing phenomenon was subsequently observed  by all LHC central detectors and  with data at a higher energy of $\sqrt{s}=$ 13 TeV (e.g. for CMS see Ref. \cite{V. Khachatryan:2016}). 
This was a very surprising result and so far  has not been fully understood. This observation revived the 
interest in studying further potential QGP sensitive variables in nucleon-nucleon
collisions, i.e. in a much simpler collision 
systems.

%this study was also done, in 2016, on the CMS runII data at 13 TeV \cite{V. Khachatryan:2016}, the same intriguing phenomena observed again not only at high multiplicity but also at minimum bias events.  

In that context, it is of interest to have a fresh look 
if the aforementioned multiplicity 
density fluctuations 
 using the high energy and high rate 
LHC data and search for events with anomalous high 
particle density spikes. Observing such events can help to shed 
light on the origin of both the lower energy observations and
the unexpected correlations measured in high energy $pp$
collisions.

This short paper reports on the preparation for such a 
multiplicity fluctuations study  
using the LHC data, deploying the most general
Monte Carlo event generators presently used for generating minimum bias $pp$ collisions
at the LHC. A minimum bias events is a technical term for a 
class of selected (triggered) collisions with an experimental
on-line trigger that has a minimal selection bias
on the collected data samples. 
Typically more than 95\% of all  
produced inelastic events pass a minimum bias 
trigger at the LHC, but due to the extremely high production rate, only a very small fraction
of these can be recorded and stored for off-line analysis, using the so-called trigger 
pre-scale factors. The special interest are on events with a very
high charged-particle multiplicity, typically multiplicities 
larger than 100, which have a much lower production rate, and can be recorded without such
pre-scale factors by the experiments.

The studies reported in this paper intend to show the characteristics of the data expected at the LHC, as predicted by the Monte Carlo event generators. 
These Monte Carlo generators are based 
on phenomenological models containing both
soft low-$p_T$ (transverse momentum) peripheral scattering collisions, as well as Standard Model QCD and electro-weak 
hard scattering processes. They do not
contain a 
new physics component that leads
to anomalous high-density events and can, therefore, be treated as a baseline estimate of the expected data in the absence of any such new
effects.

In this paper, we studied events with a high-density of charged particles within  pseudorapidity windows of different widths taken as $\Delta\eta = 0.1$, 0.2, and  0.5. 
The resolution in $\eta$ for e.g. the central tracking detector of the CMS experiment is much better, by a factor of 3-5 than the most narrow
$\Delta\eta$ interval we consider for the analysis here.

The multiplicity dependence  of the average of the maximum number of charged particles inside the selected window has been studied as well. The  PYTHIA and HERWIG Monte Carlo programs 
have been  used to simulate the $pp$ collisions in a wide range of energies 
at the LHC. Two samples are considered, a full minimum bias sample and a selected high multiplicity sample.
Note that, the central experiments at the LHC, such as ATLAS and CMS,
do not have full coverage of the particle production but, for charged tracks, only cover the central part $|\eta|<2.5$, which will be emulated in the study.

\section{Analysis}

\subsection{PYTHIA and HERWIG event generators}

%Give a short description of Herwig and Pyhtia

PYTHIA \cite{PythiaPhysicManual} and HERWIG \cite{HerwigPhysicsAndManual} are  general-purpose Monte Carlo event generators that are actively  used at the LHC. These generators have undergone decades of development and tuning to collider and other data.
%from the early version that were written in FORTRAN to the current one based on C++ programming language \cite {bucklyGeneralPurpose}.

The event generation for both generators consists of 
 several steps starting typically from a hard scattering process, followed by initial- and final-state parton showering, multi-parton interactions, and  the 
 final hadronization process. HERWIG and PYTHIA use different model approaches for these steps, e.g. PYTHIA uses a $p_{T}$-ordered approach \cite{pythiapartonshowering} for modeling of parton shower while HERWIG deploys angular-ordered showers \cite{herwigpartonshowering}. Multi-parton scattering in PYTHIA uses the original impact parameter model \cite{mulitpleinteractionpythia} while HERWIG applies the eikonal multiple partonic scattering model \cite{mulitpleinteractionherwig}. The last step is the hadronization (fragmentation) of partons into 
 hadrons; this process is performed in PYTHIA with the Lund string fragmentation model \cite{pythiahadronization1,pythiahadronization2} and an alternative model used by HERWIG is based on
 cluster fragmentation \cite{herwighadronization}. Both Monte Carlo 
 programs include a model to generate soft hadronic 
 collisions which make up the largest part of a sample of minimum bias events.  Note also that, the minimum bias events contain not only soft events but a mix of all event classes produced in the $pp$ interactions according to their 
relative cross-sections.

%\subsection{MC generator details}

The proton-proton collisions are generated using the HERWIG 7.1.5 \cite{herwig}  
and the PYTHIA 8.186  \cite{pythia} versions of the programs. The collisions are generated at  $\sqrt{s} = 2.36$, $7$, $8$, and $13$ TeV corresponding to the energies at which the LHC was operated from 2010 till 2018. About 5.10$^6$  
events were generated for both HERWIG and PYTHIA at each C.M. energy using the minimum bias generation settings of the generators. 

For  PYTHIA, the inelastic (diffractive and non-diffractive) proton-proton collisions were simulated by using the PYTHIA Monash 2013 tune \cite{monsh}. The Monash parameters are tuned in such a way that these  provide a good description of the experimental data at the LHC energies e.g. the minimum bias charged multiplicity and other event characteristics.

For HERWIG the minimum bias events were simulated by using the baryonic reconnection tune  which is a set of parameters recommended for usage when running the baryonic color reconnection model for  LHC energies. It was tuned to
generate minimum bias LHC data at $\sqrt{s}= $ 7 TeV and $\sqrt{s}= $ 13 TeV, and provides  a good description of charged-particle multiplicities and particle flavor observables in $pp$ collisions \cite{baryonproduction}. The values of baryonic reconnection parameters are available in Ref.\cite{param}.

The events and particles are selected for this analysis according to the following criteria. For the minimum bias selection, each event must have at least one charged particle in the final-state which emitted within the pseudorapidity range $|\eta| <$ 2.5
(in the full acceptance of the azimuthal angle  $\phi$) and with minimum $p_T$ $>$ of 100 MeV.
Experiments such as UA5 (no magnetic field)
and NA22/23 (bubble chambers) have negligible
particle losses at low momenta, but in an experiment such as
CMS, with a very strong magnetic field, the 
lower limit on the $p_T$ of a particle to
be reconstructed with sufficient quality
is about 100 MeV\cite{Khachatryan:2010nk}.
The effect on this measurement is however
very small, as will be shown later.

%was applied to emulate the range of the detector coverage. 
Furthermore, the high multiplicity event samples were selected with a multiplicity cut on the charged
particles, depending on
the center of mass energy of the collisions: events with a minimum number of produced charged particles larger than 100 for $\sqrt{s}=$ 13 TeV, larger than 80 for 7-8 TeV and larger than 70 for 2.36 TeV. These selections correspond
to realistic multiplicity thresholds as 
have been applied at the trigger level in the CMS experiment. The number of events used, which pass the multiplicity cuts, are given in table 1.

\begin{table}[h]
\centering
\small
\caption{The number of high multiplicity events after cutting at different C.M. energies.}
{\begin{tabular}{@{}cccc@{}} \toprule
$\sqrt{s}$ $(TeV)$ & Multiplicity Cut & HERWIG &PYTHIA \\
\hline
13 &$>100$&163787 &297427 \\
8  &$> 80$& 237936 & 364641\\
7  &$> 80$& 211851 & 320381\\
2.36  &$> 70$& 93735 & 136225\\
\hline

\end{tabular}}
\end{table}

\subsection{Maximum particle density method and the relation between its average and multiplicity}
For each event, the maximum number of charged particles 
$n_{ch}$ within
a window of pseudorapidity, of say $\Delta\eta$ = 0.1,  was calculated  by sliding this window over the full pseudorapidity range, each time the window is starting at 
an $\eta$ value of a particle in the event and ending at $\eta$ $- 0.1$. The sliding window found with the maximum number of particles is used for filling the histograms.

The same distribution was made for $\Delta\eta$ = 0.2 and $\Delta\eta$ = 0.5 windows and for all different collision energies. This method was employed both with the full minimum bias event sample and with the high multiplicity selected events.

 The average of the maximum density  of the emitted charged particles $<(\Delta n/ \Delta\eta)_{max}>$ as a function of the event multiplicity measured within the  pseudorapidity range $|\eta|<2.5$ is reported  as well.

\section{Results and Discussion}

Figs. 1.(a,c,e)  show the distribution for minimum bias events of the maximum
charged-particle density found within $\Delta\eta$ = 0.1, 0.2 and 0.5, calculated with PYTHIA and HERWIG for collisions at C.M. energies of 2.36 and 7 TeV. Figs. 1.(b,d,f) show the same distributions for collision energies  8 and 13 TeV. All figures show,  for a minimum of two particles, the expected number of events is decreasing with increasing of $n_{ch}$ for $n_{ch} > $ 5, and the width of the distributions increases with increasing of the  C.M. energy: for higher energies larger values of $n_{ch}$ are reached, as expected. For all energies, the relative
number of events with large density  produced by  PYTHIA is slightly higher than the 
events produced by HERWIG.
In fact, the ratio of the PYTHIA to HERWIG $dN/dn_{ch}$ distribution shows that the ratio is fairly constant up to $n_{ch}$ is 5 (10)
for the search with $\Delta\eta$ = 
0.1 (0.5) but rises to a value of 2-4 at the 
highest densities. This is not unexpected and, 
in most part, due to the different hadronization model used in these two Monte Carlo programs.

Similar distributions for high multiplicity selected
events are shown in Figs. 2.(a,c,e) for PYTHIA and HERWIG at $\sqrt{s}= $2.36 and 7 TeV respectively, and Figs. 2.(b,d,f) are for  $\sqrt{s}= $ 8 and 13 TeV.

It is interesting to notice that the effect of the
100 MeV $p_T$ cut-off in the collider experiments does not have a significant impact on this measurement as expected by PYTHIA and still small but slightly larger as expected by HERWIG. The influence of the $p_T$ cut predicted by  PYTHIA is shown  in Fig. 3,  which compares the distributions with and without  the $p_T$ $>$ 100 MeV cut,  at C.M. energy 
of 13 TeV and $\Delta \eta =$ 0.1.

The rate of events with a large density of charged particles inside a small window $dN/dn_{ch}$ decreased exponentially with increasing of the maximum number of charged particles inside the same window size $n_{ch}$, according to the relation $dN/dn_{ch} = a$ $exp(-bn_{ch})$ as suggested also by the measurements in earlier experiments. Fitted values of the parameters $a$ and $b$
 are given in table 2.

Generally, this gives good fits of the data, with a $\chi^2$/NDF around one, showing a smooth single component for the shape of  these distributions.
The b-parameter is physics-wise the most significant one and allows
e.g. to compare with measurements from earlier
experiments. 
It is interesting to note that we find the value of 
the b-parameter for the 
search with $\Delta\eta= 0.1$, is around
one and 
essentially identical to the values found 
and reported in \cite{Adamus:1986vd} for 
similar fits. 
Exponential fits were also made on the
minimum bias distributions. The b-parameters
were found to be very similar, generally 
larger by a few  \% compared to those  derived from the high multiplicity
distributions hence showing that the bias
for the high multiplicity selection is very small, and will give a representative
account for the anomaly search.

Clearly, and as expected,
we observe a few events at high $n_{ch}$ 
values where less than one event would be 
expected based on these fits, due to statistical fluctuations in the data,
but we do not observe a significant tail.
These fluctuations can lead to single events that 
can pass the maximum expected $n_{ch}$ value from the fits, extending the $n_{ch}$ range
 by approximately 20-30\%.
Hence, the conclusions of future real measurements made with the
LHC data cannot be based just on a few events outside the 
expected region.

For the analysis of the $\Delta \eta = 0.1$ selection, we find four outlying events. Three
of these events are visible in Figs. 1 (a,b) with 20, 25, and 26 particles in this small 
pseudo-rapidity interval. This corresponds to 
an expectation  probability to observe such an event of 0.01-0.005
calculated  using the exponential fit 
predictions given in 
table 2. One very anomalous event -not shown in the figures-  
was produced in the PYTHIA  generation at a C.M. 
energy of 8 TeV, and gave as much as 48
charged particles in an interval $\Delta \eta = 0.1$! This event has a total charged multiplicity of 464 which
is exceptionally high. It was analyzed in
detail and found to be a  QCD multi-jet 
event with the characteristic of having four multi-parton interactions (MPIs).
This demonstrates again 
that a low number of observed anomalous events will 
always be difficult to be conclusive.

If we further specifically search for signals of models
related to a QCD Cherenkov radiation, a 
relative uniform distribution of the particles 
in the $\phi$ variable is expected. 
We show for the four
 “anomalous events" discussed above an event 
display of the particle distribution in the 
$\phi-\eta$ plane and indicate on the figure
the interval found with the highest density
of particles, see Fig. 4. We, clearly, observe that the particles strongly cluster in the  $\phi-\eta$ plane, likely caused by QCD radiation and jetiness, unlike what would be expected from such 
a new QCD Cherenkov effect. Hence the 
simultaneous analysis of the  $\Delta \eta $ outlying events with high densities and the $\phi-\eta$ plane analysis should help
to establish the genuine anomalous nature of the 
events.

The relation between $<(\Delta n / \Delta\eta)_{max}>$ and 
the total charged-particle multiplicity in the region of $|\eta| <2.5$ is represented in  Figs. 5.(a,c,e) for PYTHIA and Figs. 5.(b,d,f) for HERWIG. Fig. 5 shows an almost linear dependence of $<(\Delta n / \Delta\eta)_{max}>$ on the multiplicity for all pseudorapidity  window sizes and all  energies. This relation is also independent on the energy up to a multiplicity of  120 which  was also
observed by the UA5 and NA22 results.  For  larger values than 120, the fluctuations are increased, due to the small number of
produced events with a large multiplicity.

\clearpage

\begin{figure}[!tbp]
 \begin{subfigure}{0.49\textwidth}
    \includegraphics[width=\textwidth , height=5.8cm]{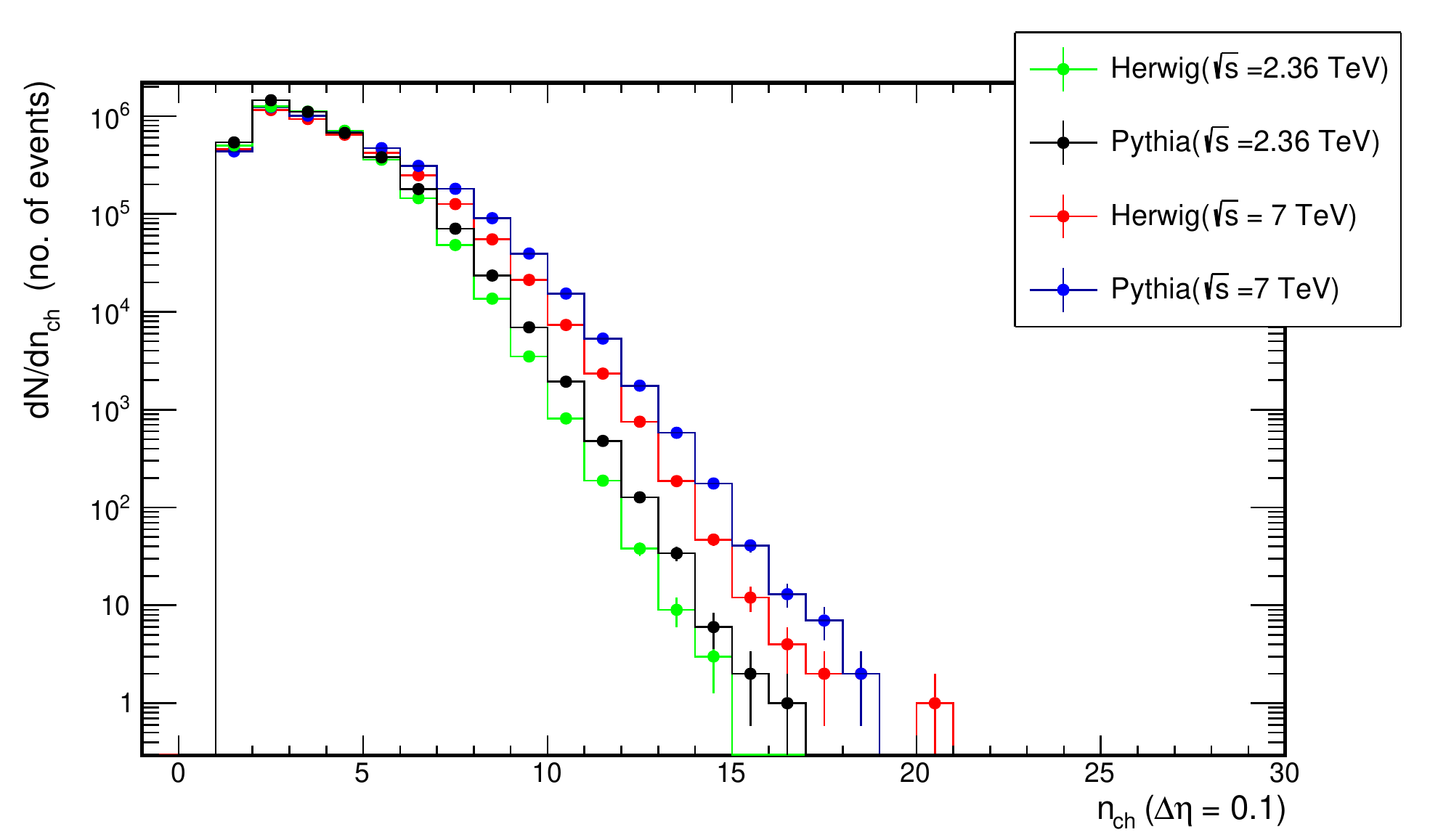}
\caption{\footnotesize for $\sqrt{s} = 2.36, 7 $ TeV at $\Delta\eta = 0.1 $ } 
\end{subfigure}
  \hfill
   \begin{subfigure}{0.49\textwidth}
    \includegraphics[width=\textwidth , height=5.8cm]{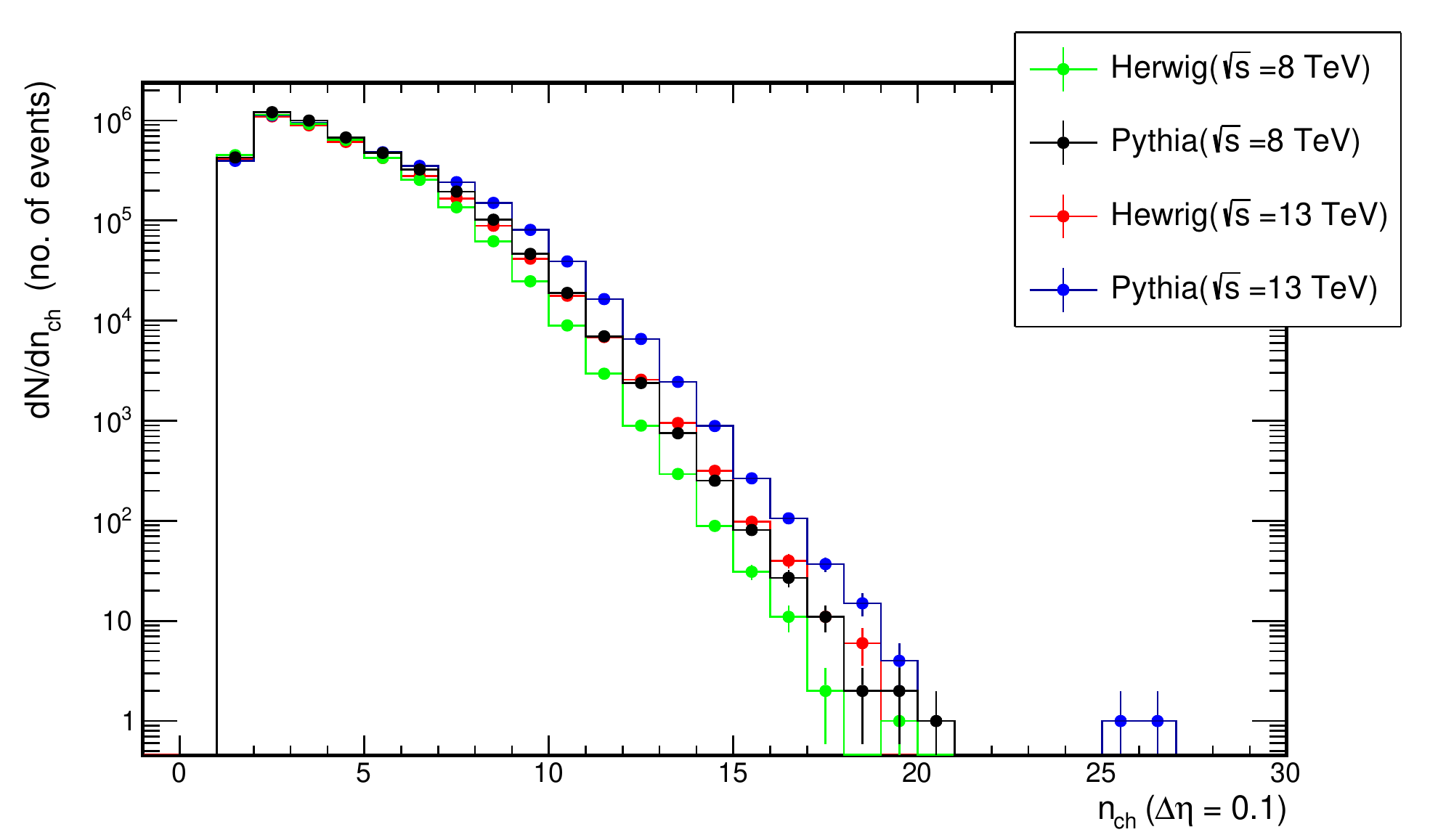}
\caption{\footnotesize  for $\sqrt{s} = 8, 13$ TeV at $\Delta\eta = 0.1 $ } 
\end{subfigure}

%\begin{figure}[!tbp]
 \begin{subfigure}{0.49\textwidth}
    \includegraphics[width=\textwidth , height=5.8cm]{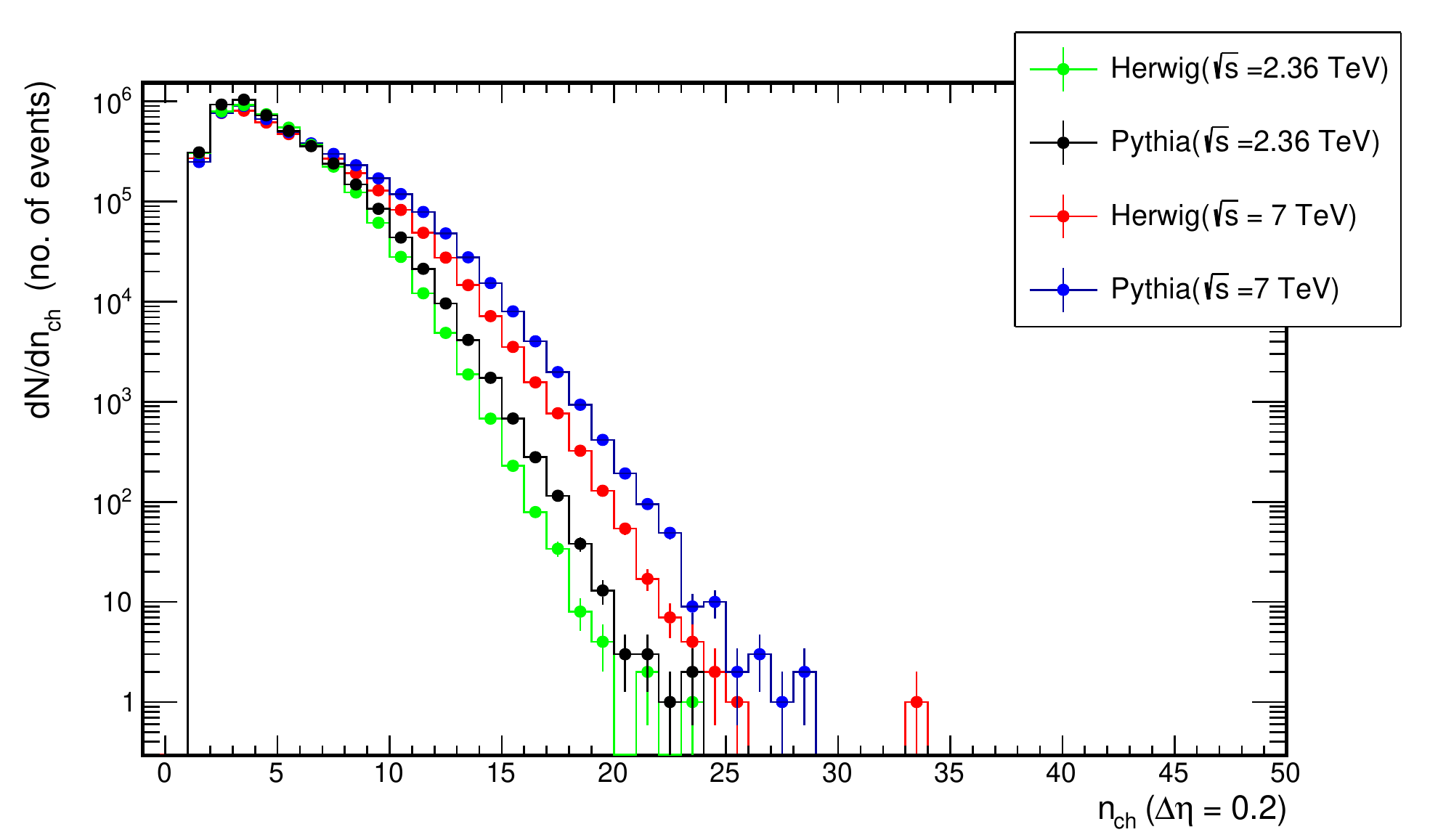}
\caption{\footnotesize for $\sqrt{s} = 2.36, 7 $ TeV at $\Delta\eta = 0.2 $ } 
\end{subfigure}
  \hfill
   \begin{subfigure}{0.49\textwidth}
    \includegraphics[width=\textwidth , height=5.8cm]{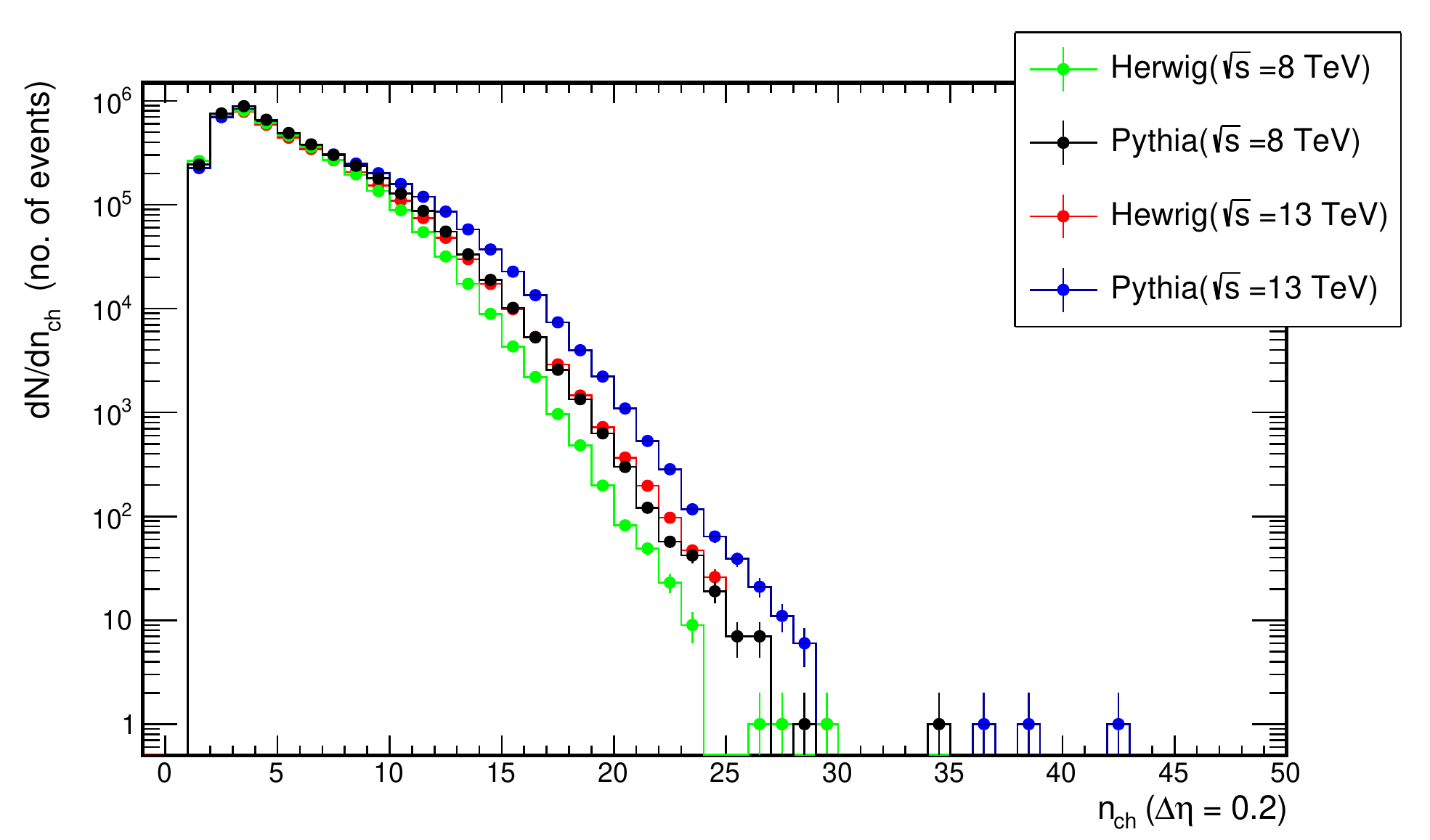}
\caption{\footnotesize for $\sqrt{s} = 8 , 13 $ TeV at $\Delta\eta = 0.2 $ } 
\end{subfigure}

%\begin{figure}[!tbp]
 \begin{subfigure}{0.49\textwidth}
    \includegraphics[width=\textwidth , height=5.8cm]{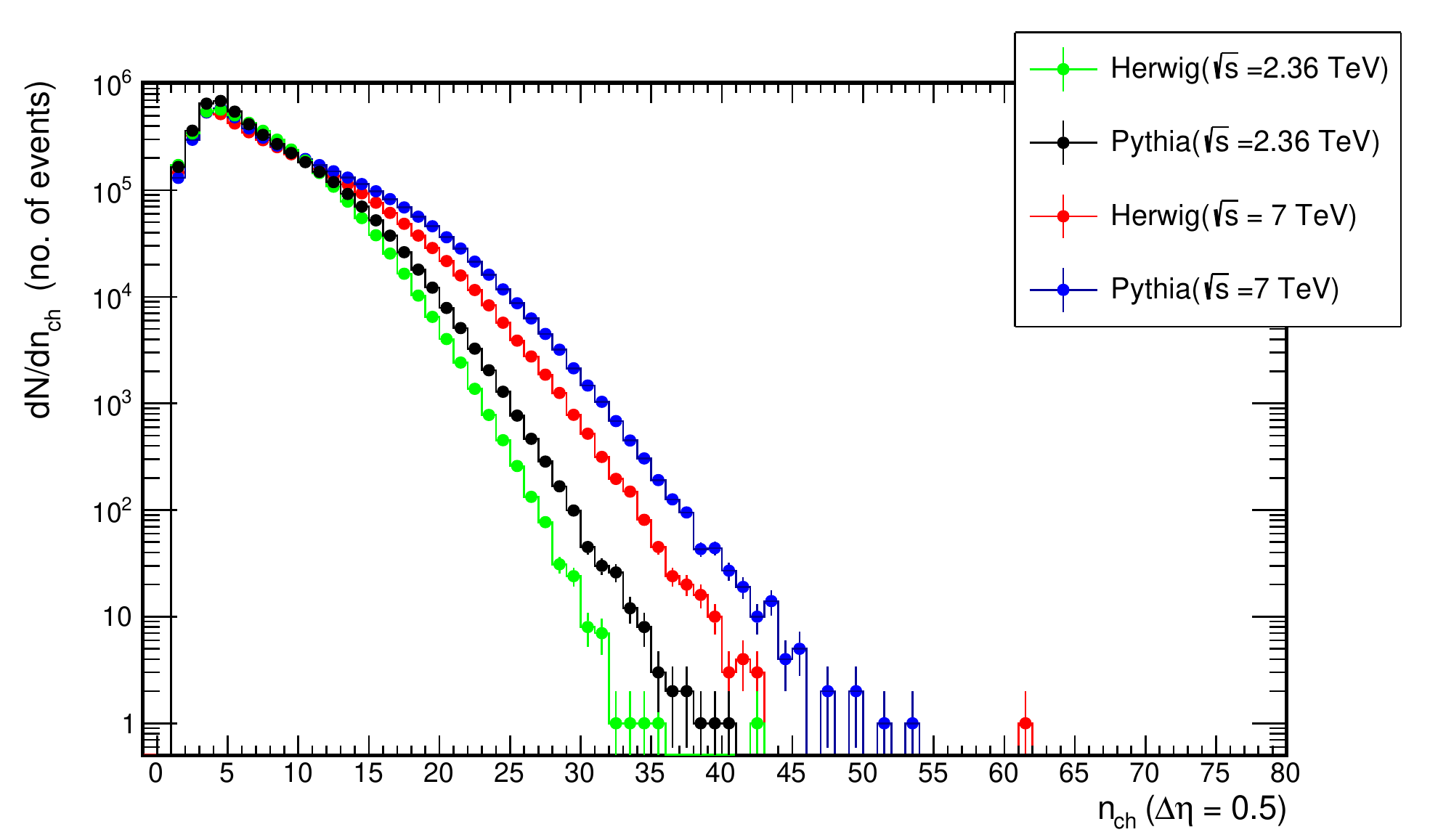}
\caption{\footnotesize for $\sqrt{s} = 2.36, 7  $ TeV at $\Delta\eta = 0.5 $ } 
\end{subfigure}
  \hfill
   \begin{subfigure}{0.49\textwidth}
    \includegraphics[width=\textwidth , height=5.8cm]{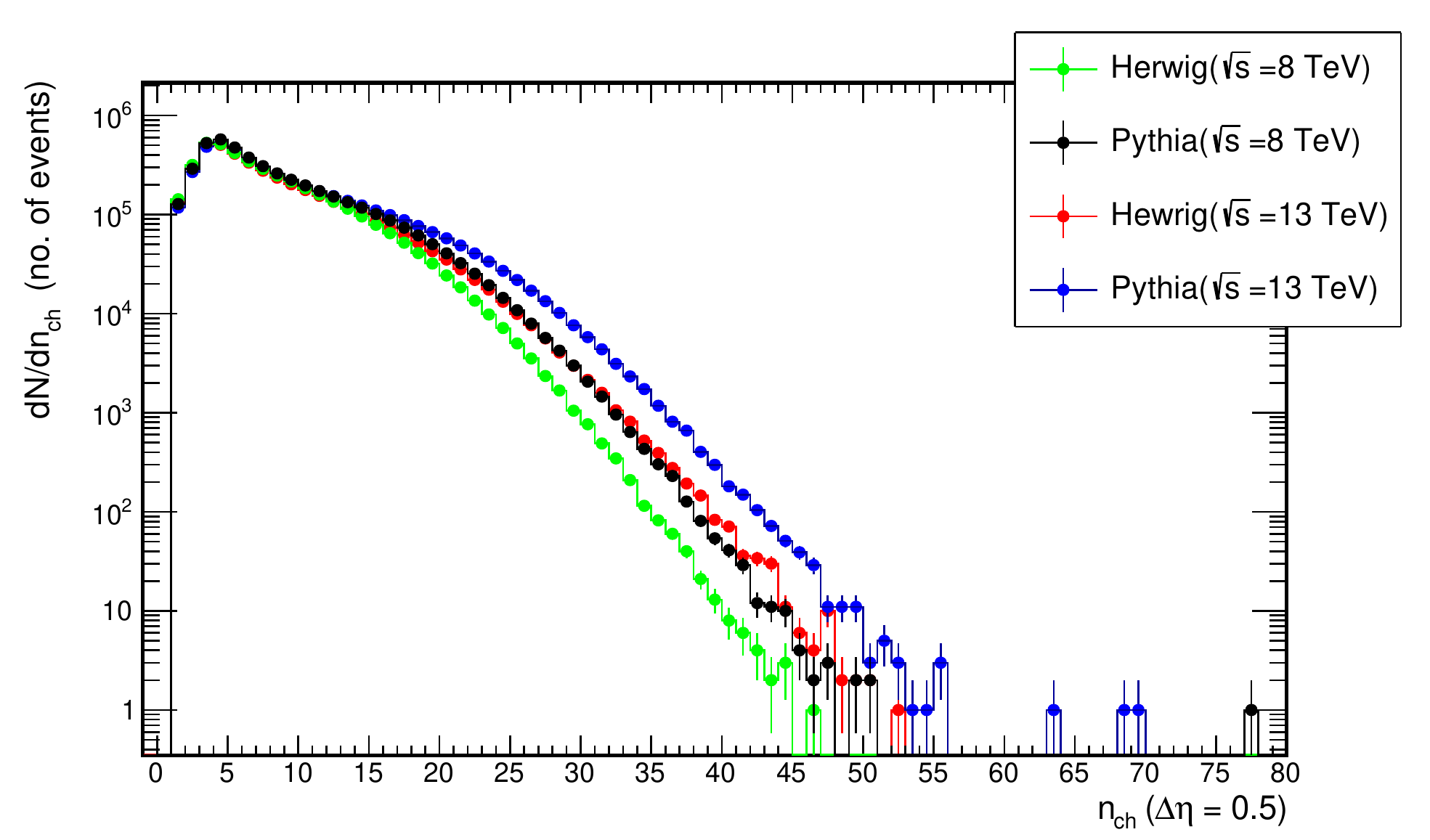}
\caption{\footnotesize for $\sqrt{s} =  8 , 13 $ TeV at $\Delta\eta = 0.5 $ } 
\end{subfigure}
\caption{ 
(a), (c) and (e) represents the comparison between the minimum bias distributions of the maximum density of charged particles generated by HERWIG and PYTHIA at $\sqrt{s} = 2.36$ TeV, $7$ TeV passing through $\Delta\eta = 0.1, 0.2 $ and $0.5$ respectively, and (b), (d) and (f) are for $\sqrt{s} = 8$ TeV, $13$ TeV.
} %\label{fig:5}
\end{figure}

\begin{figure}[!tbp]
 \begin{subfigure}{0.49\textwidth}
    \includegraphics[width=\textwidth , height=5.8cm]{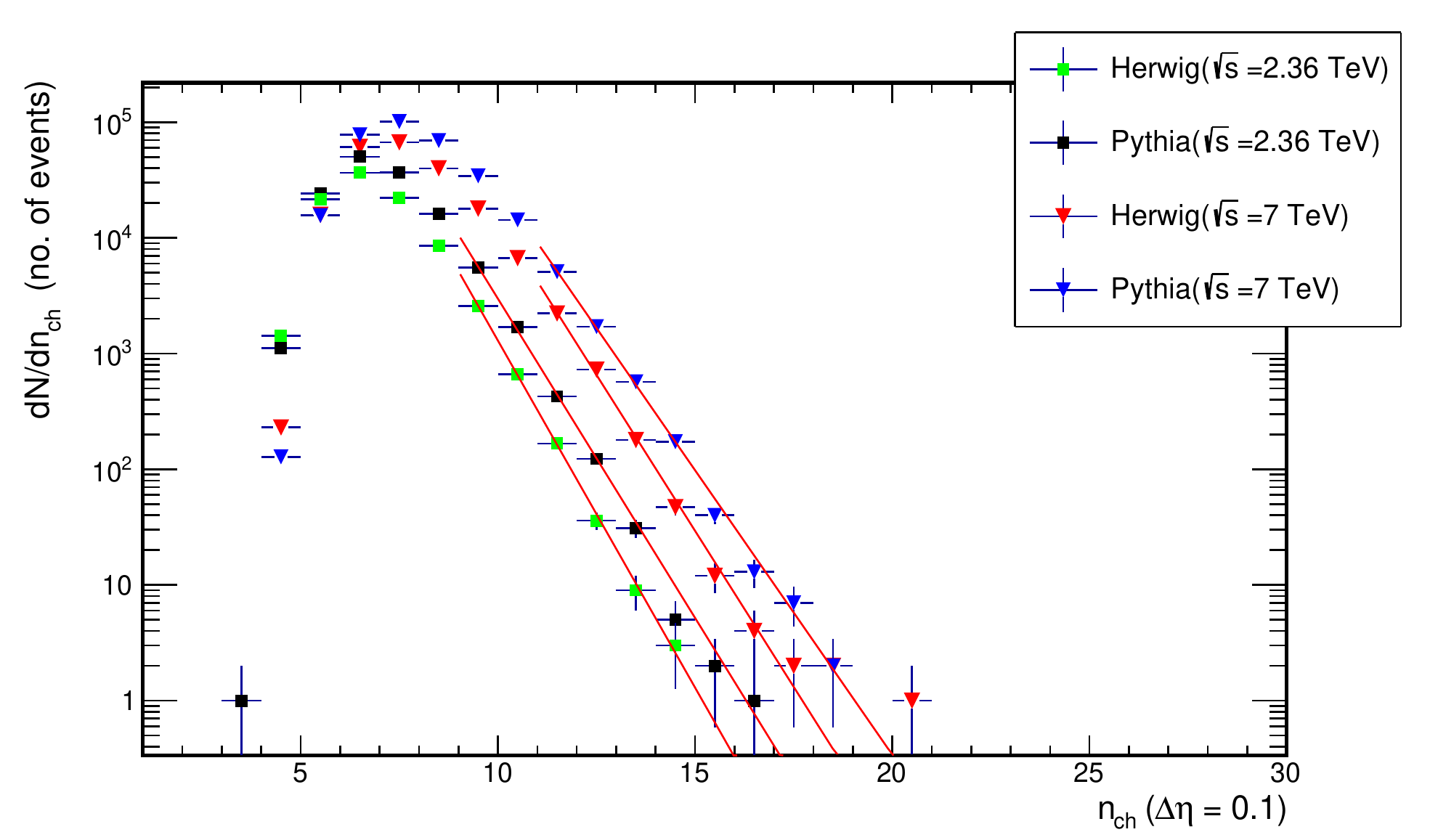}
\caption{\footnotesize for $\sqrt{s} = 2.36, 7  $ TeV at $\Delta\eta = 0.1 $ } 
\end{subfigure}
  \hfill
   \begin{subfigure}{0.49\textwidth}
    \includegraphics[width=\textwidth , height=5.8cm]{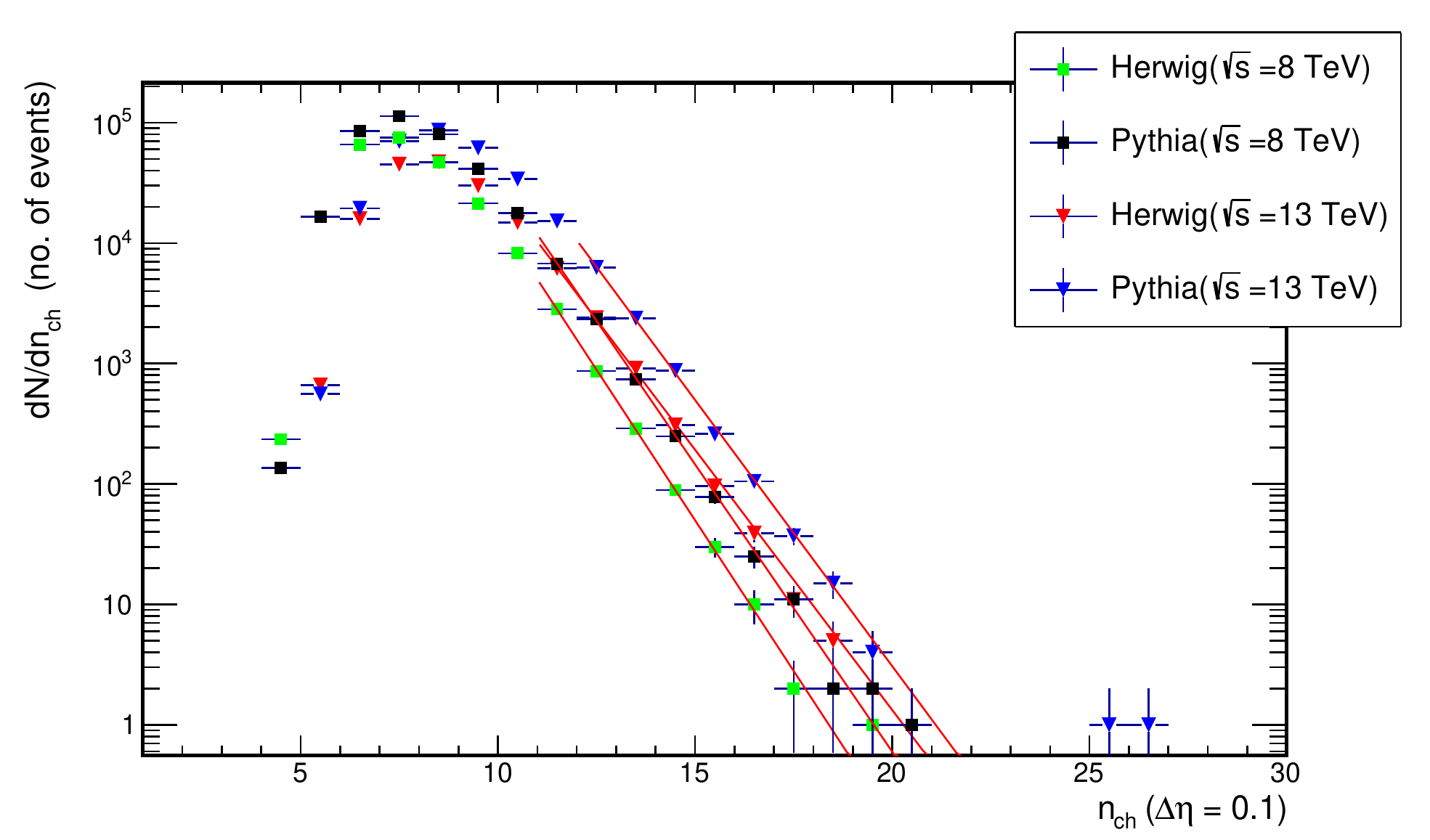}
\caption{\footnotesize  for $\sqrt{s} = 8 , 13 $ TeV at $\Delta\eta = 0.1 $ } 
\end{subfigure}

%\begin{figure}[!tbp]
 \begin{subfigure}{0.49\textwidth}
    \includegraphics[width=\textwidth , height=5.8cm]{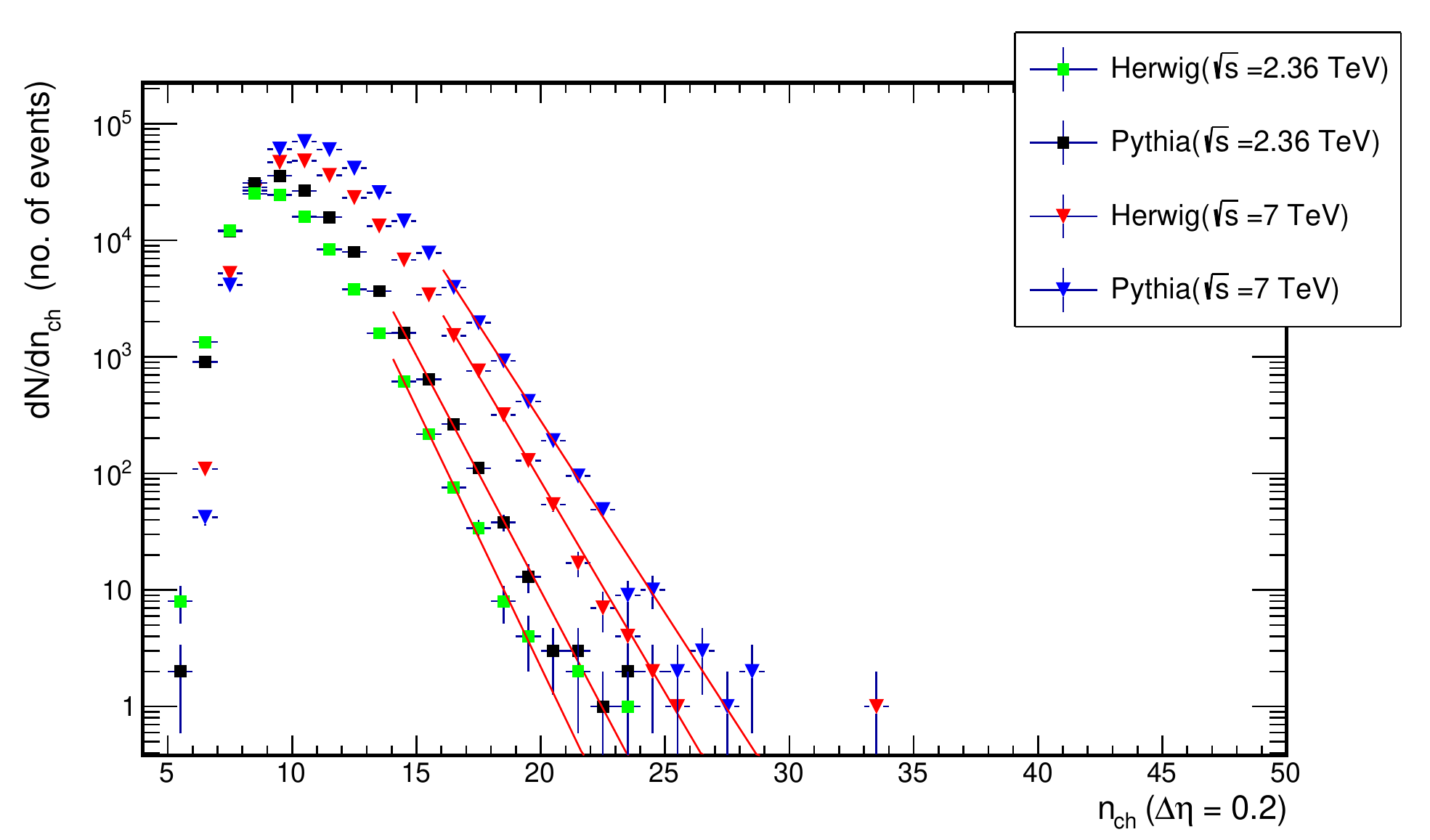}
\caption{\footnotesize for $\sqrt{s} = 2.36, 7  $ TeV at $\Delta\eta = 0.2 $ } 
\end{subfigure}
  \hfill
   \begin{subfigure}{0.49\textwidth}
    \includegraphics[width=\textwidth , height=5.8cm]{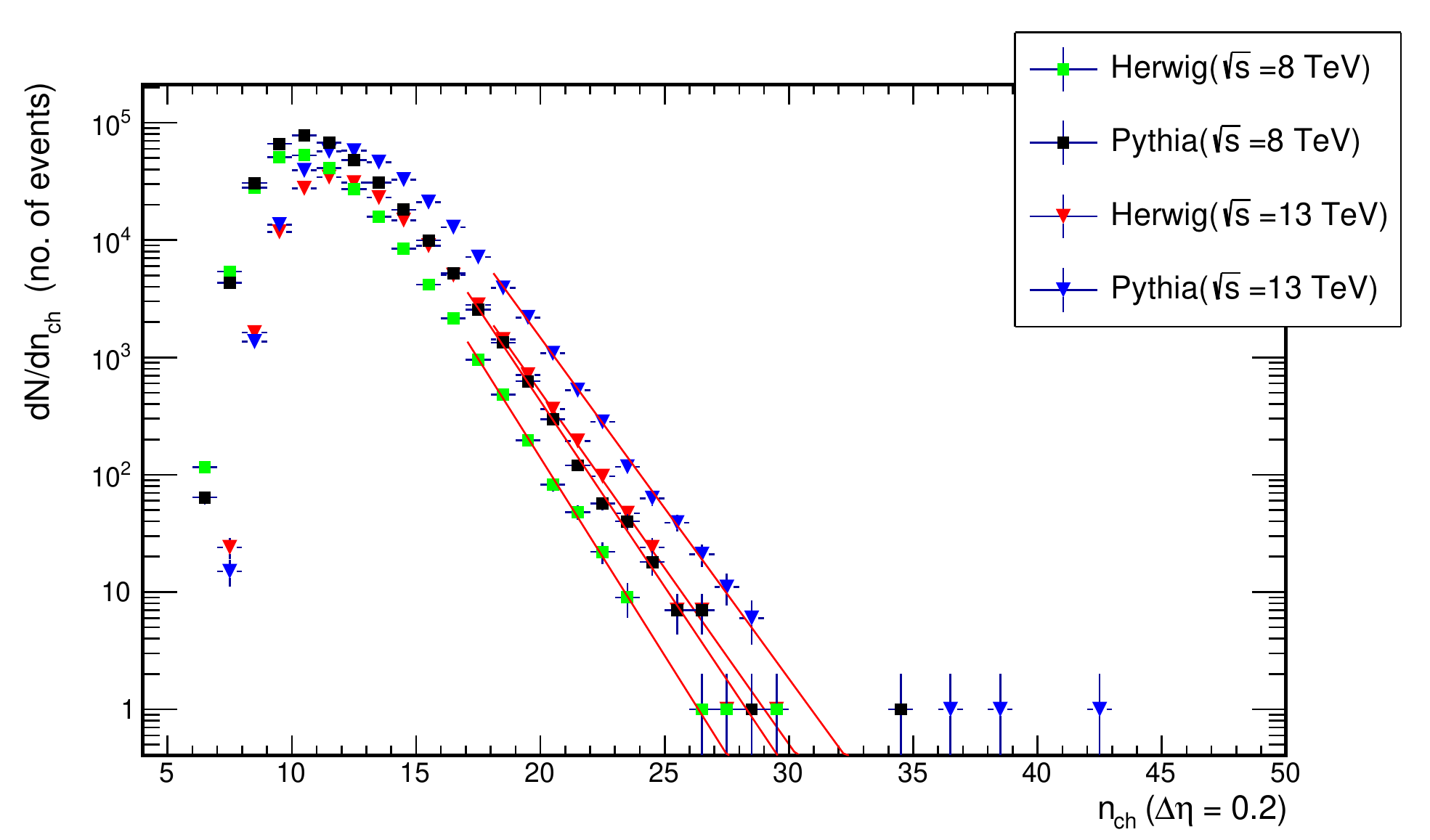}
\caption{\footnotesize for $\sqrt{s} =  8 , 13 $  TeV at $\Delta\eta = 0.2 $ } 
\end{subfigure}

%\begin{figure}[!tbp]
 \begin{subfigure}{0.49\textwidth}
    \includegraphics[width=\textwidth , height=5.8cm]{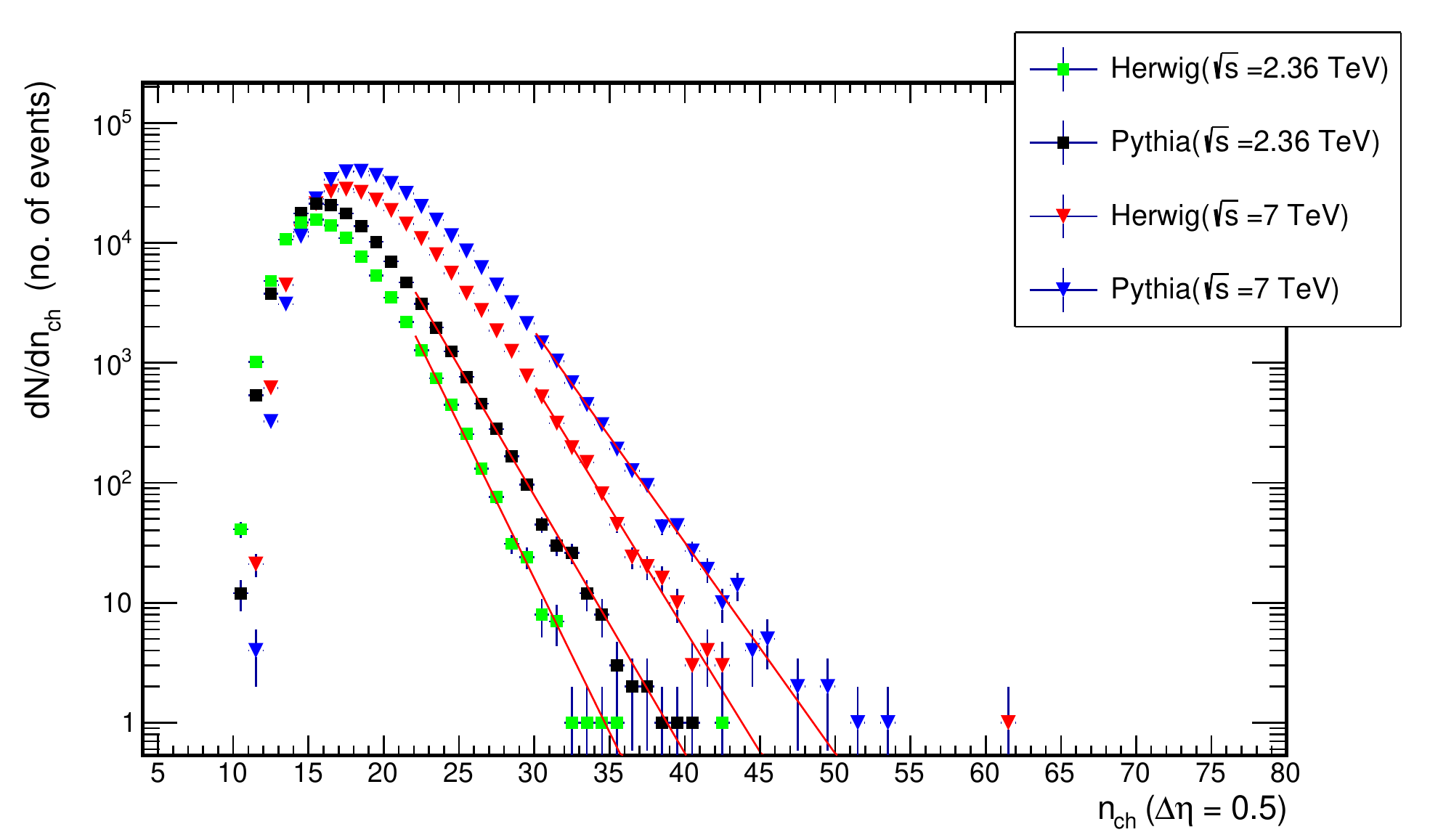}
\caption{\footnotesize for $\sqrt{s} = 2.36, 7  $ TeV at $\Delta\eta = 0.5 $ } 
\end{subfigure}
  \hfill
   \begin{subfigure}{0.49\textwidth}
    \includegraphics[width=\textwidth , height=5.8cm]{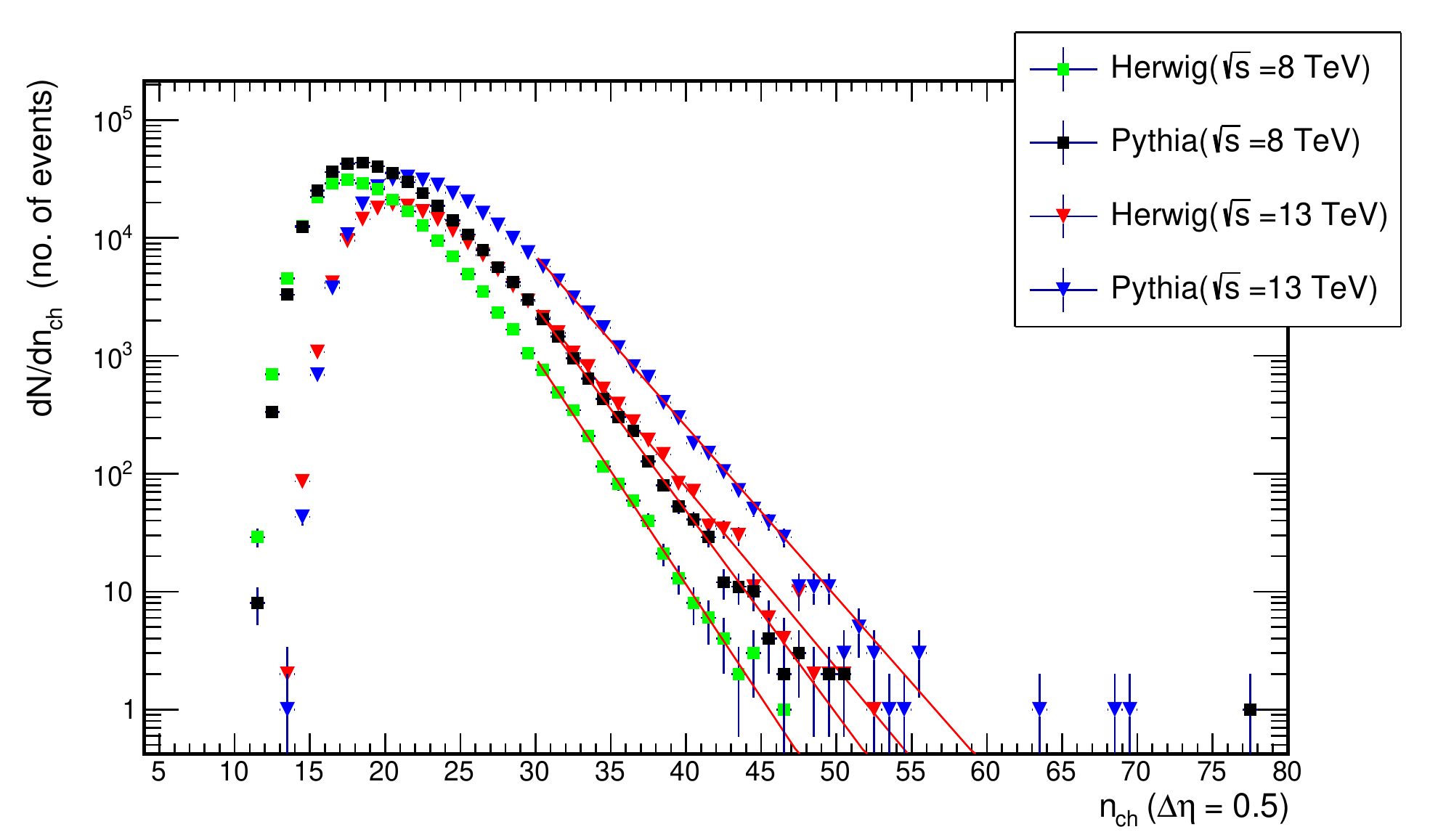}
\caption{\footnotesize for $\sqrt{s} =  8 , 13 $ TeV at $\Delta\eta = 0.5 $ } 
\end{subfigure}
\caption{(a), (c) and (e) represents the comparison between the high multiplicity distributions of the maximum density of charged particles generated by HERWIG and PYTHIA at $\sqrt{s} = 2.36$ TeV, $7$ TeV passing through $\Delta\eta = 0.1, 0.2 $ and $0.5$ respectively, and (b), (d) and (f) are for $\sqrt{s}$ = $8$ TeV, $13$ TeV.} %\label{fig:5}
\end{figure}

\begin{table}[!p]
%\small
\caption{The exponential fitting parameters of the spikes generated by  PYTHIA and HERWIG.}

  \begin{tabular}{cccc}
 
 \hline
$\sqrt{s}$ &Fitting Range & $a*10^8$ & $b$ \\
    \hline
    &&PYTHIA at $\Delta\eta$ =0.1&\\
    \hline

13 TeV    &  12-23    &$  21.40       \pm  2.90               $&$  1.018       \pm  0.010 $ \\

8 TeV     &   11-21    &$ 20.95       \pm  2.82                 $&$  1.099       \pm   0.011 $ \\

7 TeV     &   11-21    &$ 23.31        \pm  3.63                  $&$  1.132       \pm  0.013 $ \\

2.36 TeV  &   9-18     &$   9.85        \pm 1.40                $&$  1.271       \pm 0.014 $ \\

    \hline
    &&HERWIG at $\Delta\eta$ =0.1&\\
    \hline

13 TeV    &  11-23    &$   5.86       \pm  0.70               $&$  0.995       \pm  0.010  $ \\

8 TeV     &   11-21    &$ 15.99       \pm  3.70                 $&$ 1.152       \pm   0.019 $ \\

7 TeV     &   11-21    &$ 35.91        \pm  9.39                  $&$  1.242       \pm  0.022 $ \\

2.36 TeV  &   9-18     &$   12.94        \pm 3.19                $&$  1.381       \pm  0.025 $ \\

    \hline
    &&PYTHIA at $\Delta\eta$ =0.2&\\
    \hline

13 TeV    & 18-36    &$    9.64       \pm  1.40              $&$  0.670       \pm  0.074  $ \\

8 TeV     &  17-30    &$ 9.05       \pm 1.73                 $&$ 0.729       \pm   0.010  $ \\

7 TeV     &   16-30    &$ 11.14        \pm  1.62                  $&$  0.759       \pm  0.008 $ \\

2.36 TeV  &   14-26     &$   11.51        \pm 3.08                 $&$  0.929       \pm  0.018  $ \\

    \hline
    &&HERWIG at $\Delta\eta$ =0.2&\\
    \hline

13 TeV    & 18-36    &$    5.14       \pm   1.27              $&$  0.692       \pm  0.013  $ \\

8 TeV     &  17-30    &$ 7.75       \pm 2.82                 $&$ 0.777       \pm   0.020  $ \\

7 TeV     &   16-30    &$ 15.04        \pm   3.84                  $&$  0.834       \pm  0.015 $ \\

2.36 TeV  &   14-26     &$   17.18        \pm 9.32                $&$  1.024       \pm  0.036   $ \\

    \hline
    &&PYTHIA at $\Delta\eta$ =0.5&\\
    \hline

13 TeV    & 30-60    &$    1.58       \pm   0.11              $&$   0.334       \pm  0.002  $ \\

8 TeV     &  30-54   &$ 3.80       \pm0.60                $&$ 0.397       \pm   0.005  $ \\

7 TeV     &   30-52    &$3.55        \pm   0.71                  $&$  0.406       \pm   0.006 $ \\

2.36 TeV  &   22-42     &$   2.16       \pm0.27                 $&$  0.494       \pm  0.005   $ \\

   \hline
    &&HERWIG at $\Delta\eta$ =0.5&\\
    \hline

13 TeV    & 30-60    &$    1.02      \pm   0.13              $&$    0.353       \pm  0.004  $ \\

8 TeV     &  30-54   &$ 5.58       \pm 1.75                $&$ 0.442       \pm    0.010  $ \\

7 TeV     &   30-52    &$ 8.49        \pm   3.60                  $&$   0.469       \pm   0.013 $ \\

2.36 TeV  &   22-42     &$   7.71       \pm1.77                 $&$  0.590       \pm  0.010    $ \\

    \hline

  \end{tabular}
\end{table}

\begin{figure}[!tbp]

    \includegraphics[width=\textwidth , height=6cm]{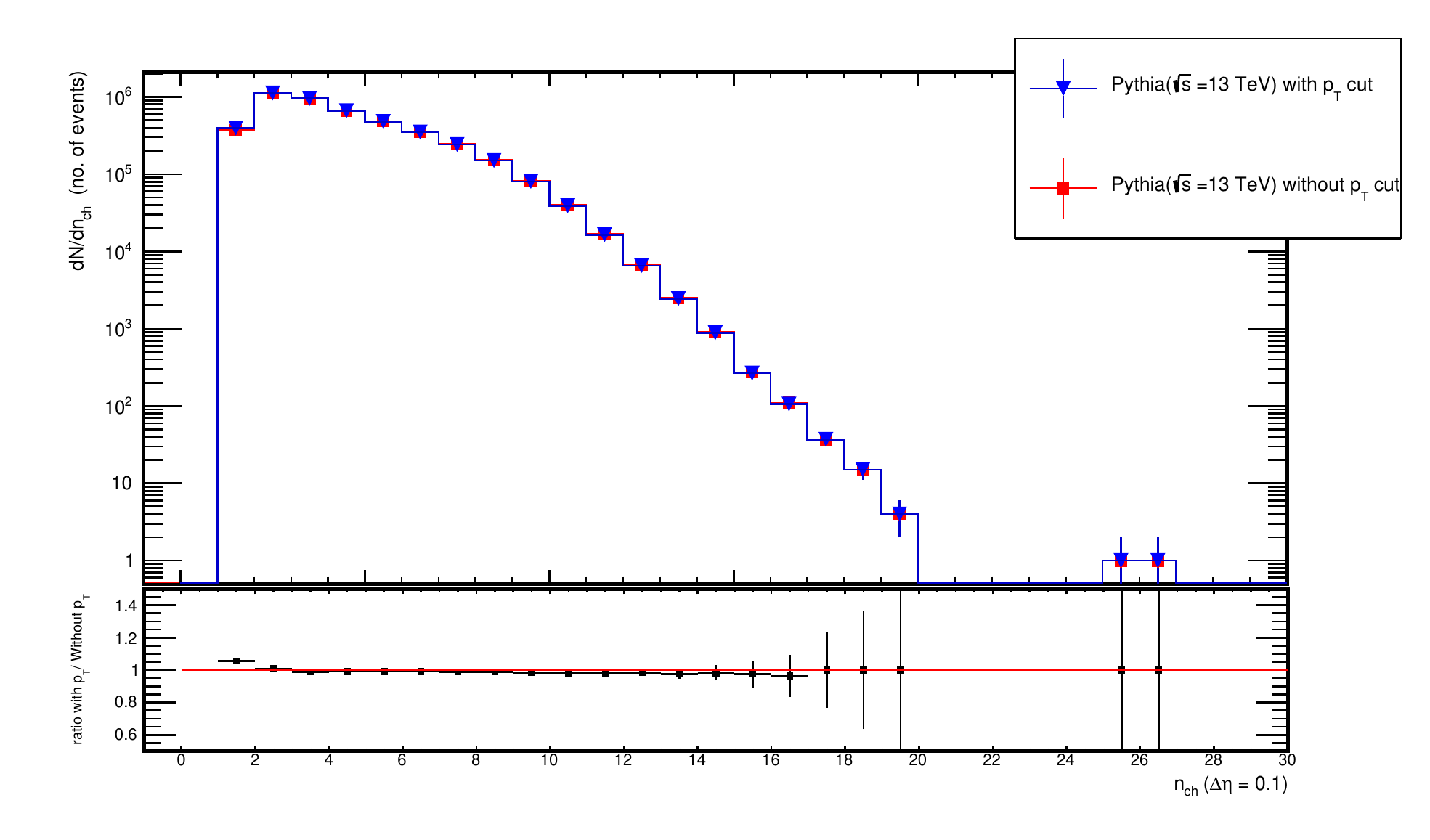}
\caption{\footnotesize Comparison between the PYTHIA minimum bias distributions (With and Without the $P_{T}$ 100 MeV cut)  at C.M. energy of 13 TeV and $\Delta\eta =0.1$.}

 \begin{subfigure}{0.49\textwidth}
    \includegraphics[width=\textwidth , height=5.5cm]{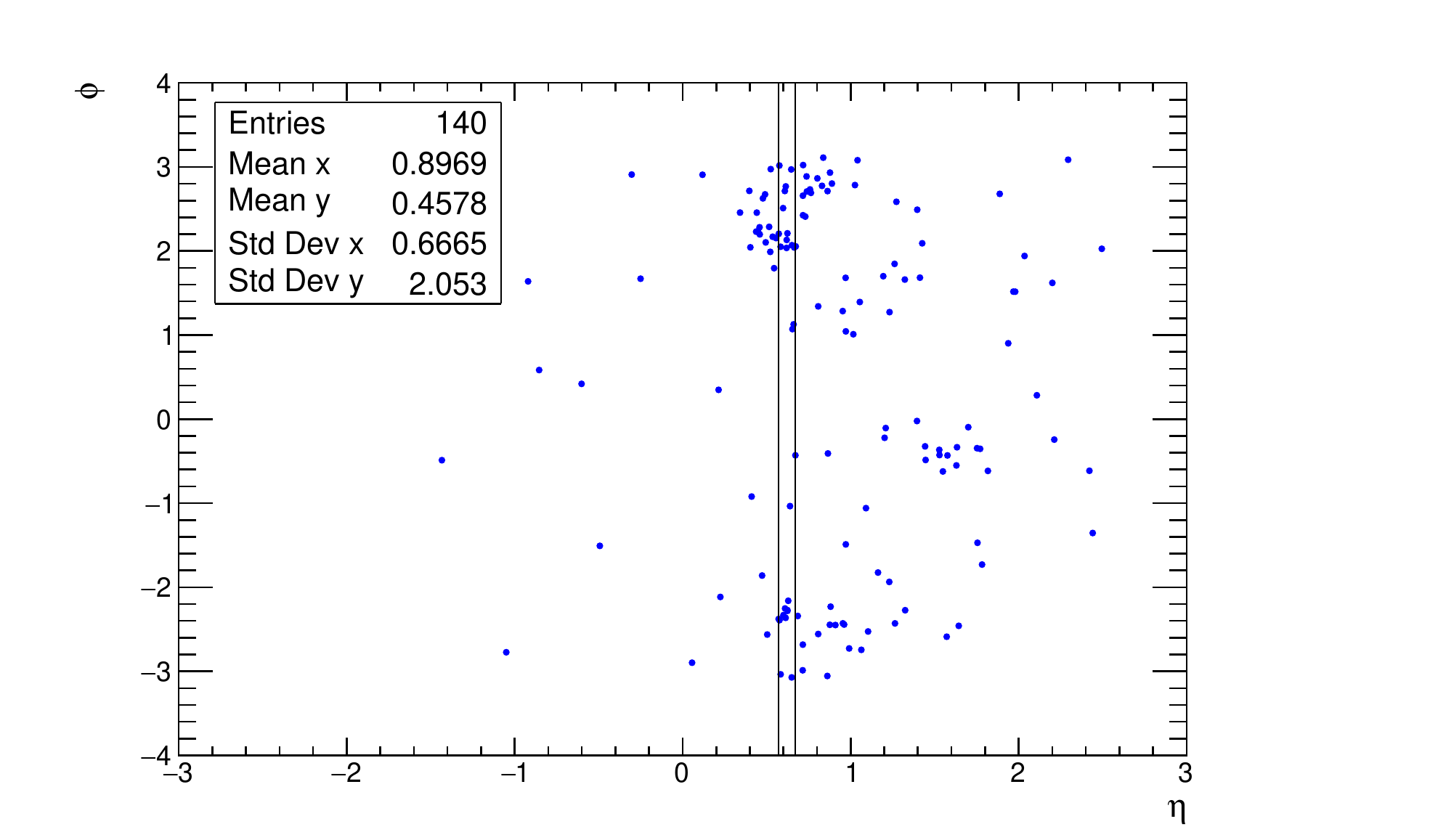}
\caption{\footnotesize for the event with 26 particles at the PYTHIA 13 TeV run } 
\end{subfigure}
  \hfill
   \begin{subfigure}{0.49\textwidth}
    \includegraphics[width=\textwidth , height=5.5cm]{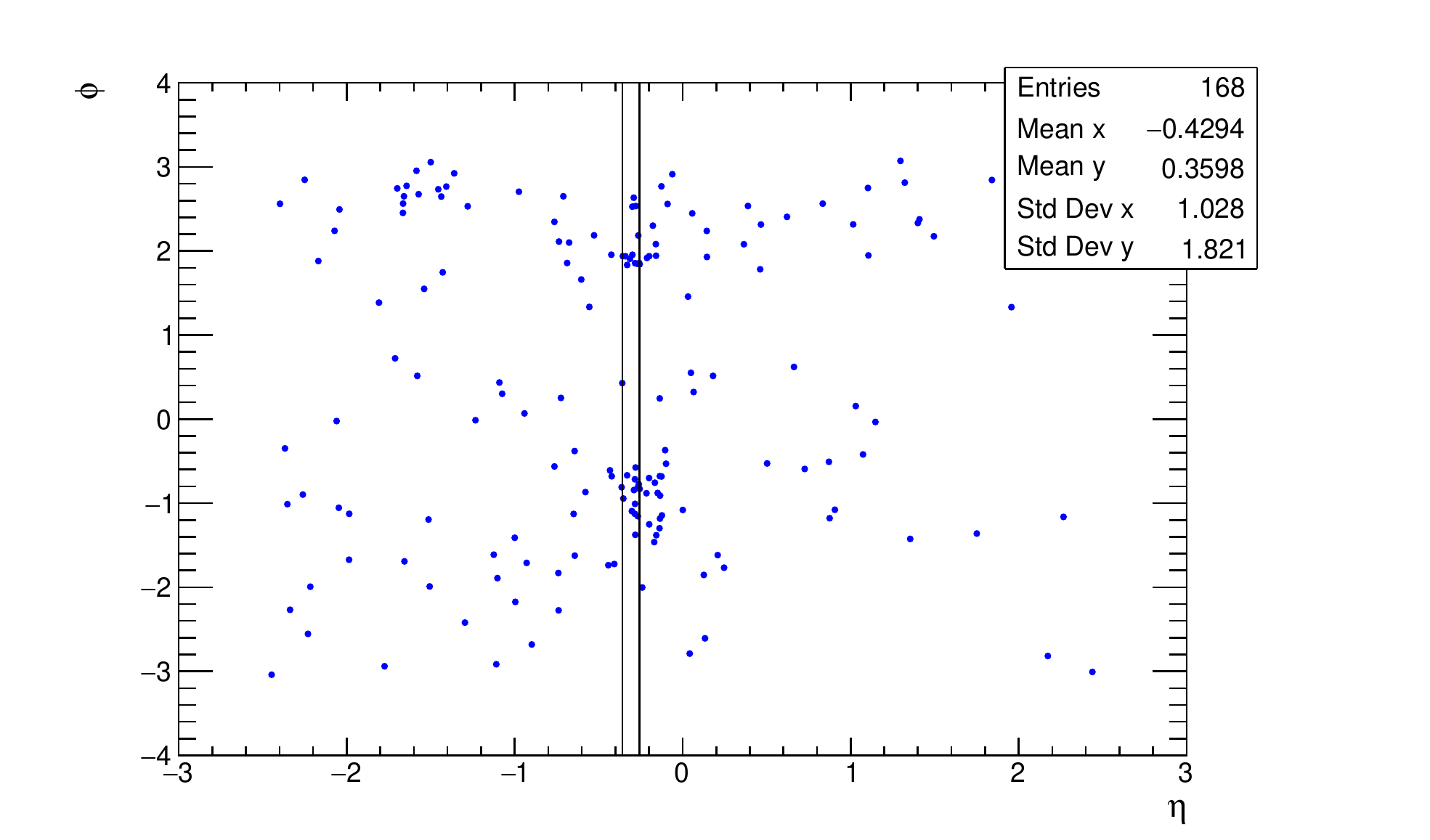}
\caption{ \footnotesize for the event with 25 particles  at the PYTHIA 13 TeV run } 
\end{subfigure}

%\begin{figure}[!tbp]
 \begin{subfigure}{0.49\textwidth}
    \includegraphics[width=\textwidth , height=5.5cm]{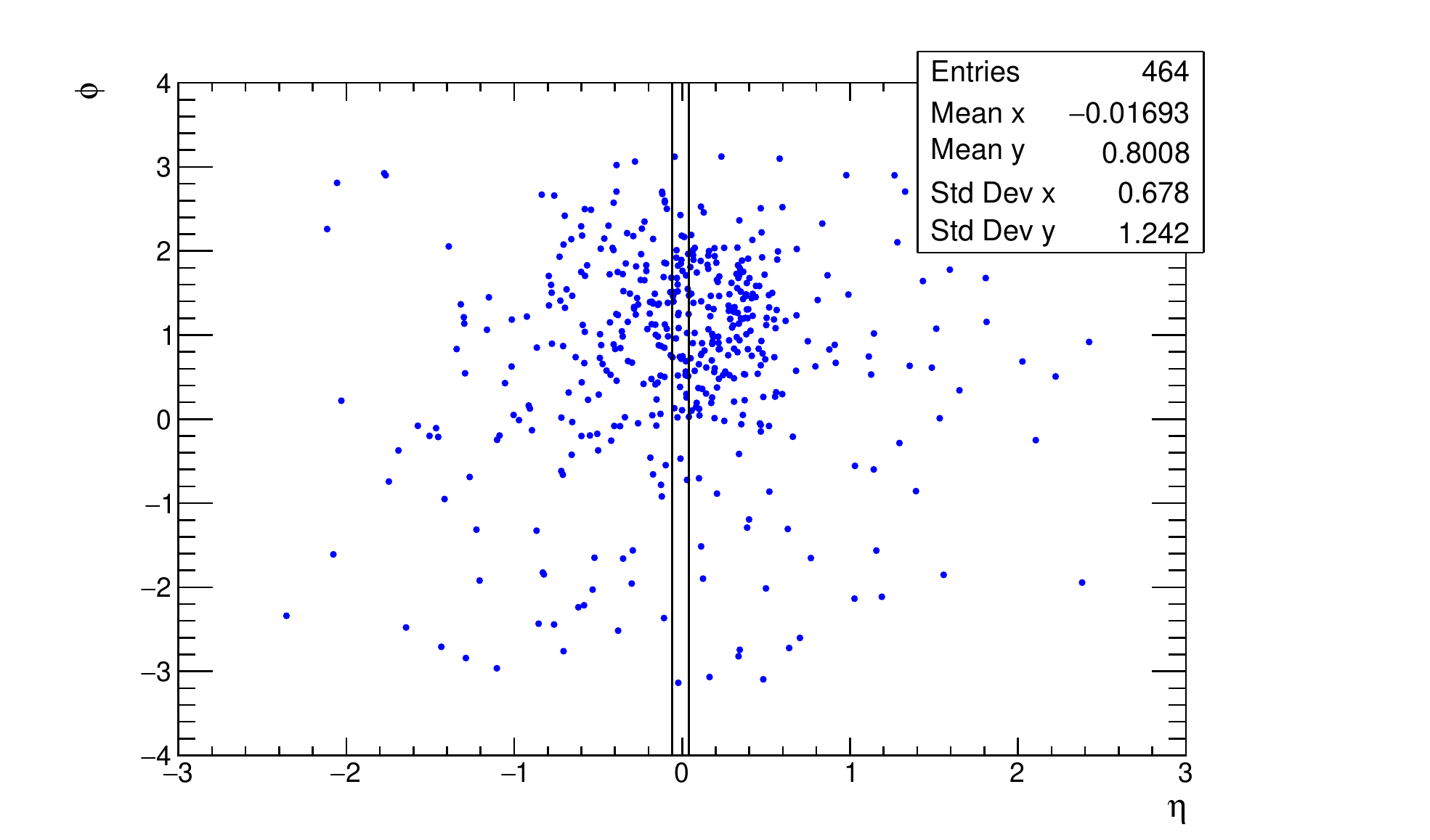}
\caption{\footnotesize  for the event with 47 particles  at the PYTHIA 8 TeV run } 
\end{subfigure}
  \hfill
   \begin{subfigure}{0.49\textwidth}
    \includegraphics[width=\textwidth , height=5.5cm]{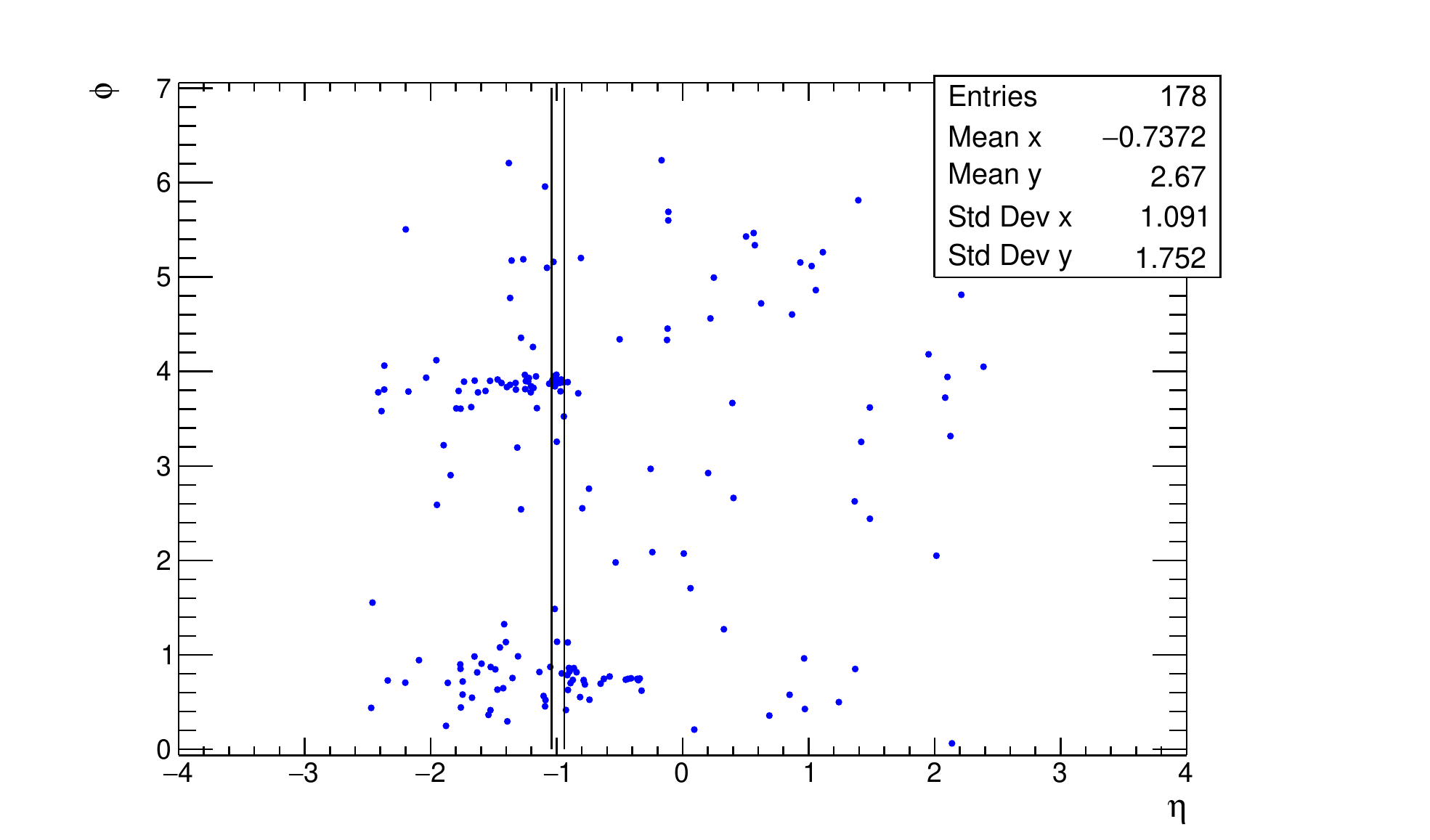}
\caption{\footnotesize for the event with 20 particles at the HERWIG 7 TeV run } 
\end{subfigure}
\caption{ The $ \phi$  - $\eta$  plane for the four outlying events at $\Delta \eta$ =0.1.} 
\end{figure}

\begin{figure}[!tbp]
 \begin{subfigure}{0.49\textwidth}
    \includegraphics[width=\textwidth , height=5.8cm]{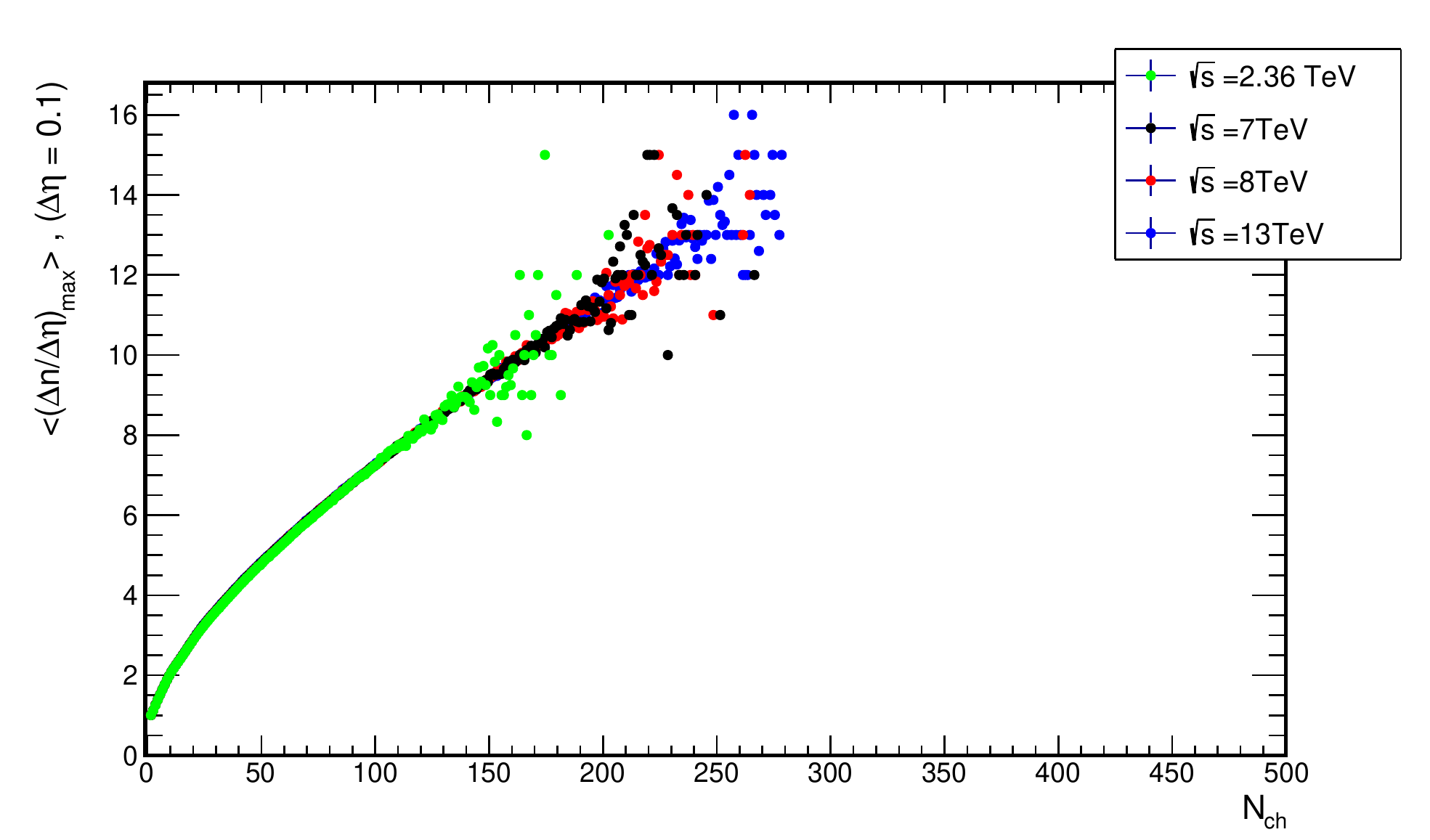}
\caption{\footnotesize for PYTHIA at $\Delta\eta = 0.1 $ } 
\end{subfigure}
  \hfill
   \begin{subfigure}{0.49\textwidth}
    \includegraphics[width=\textwidth , height=5.8cm]{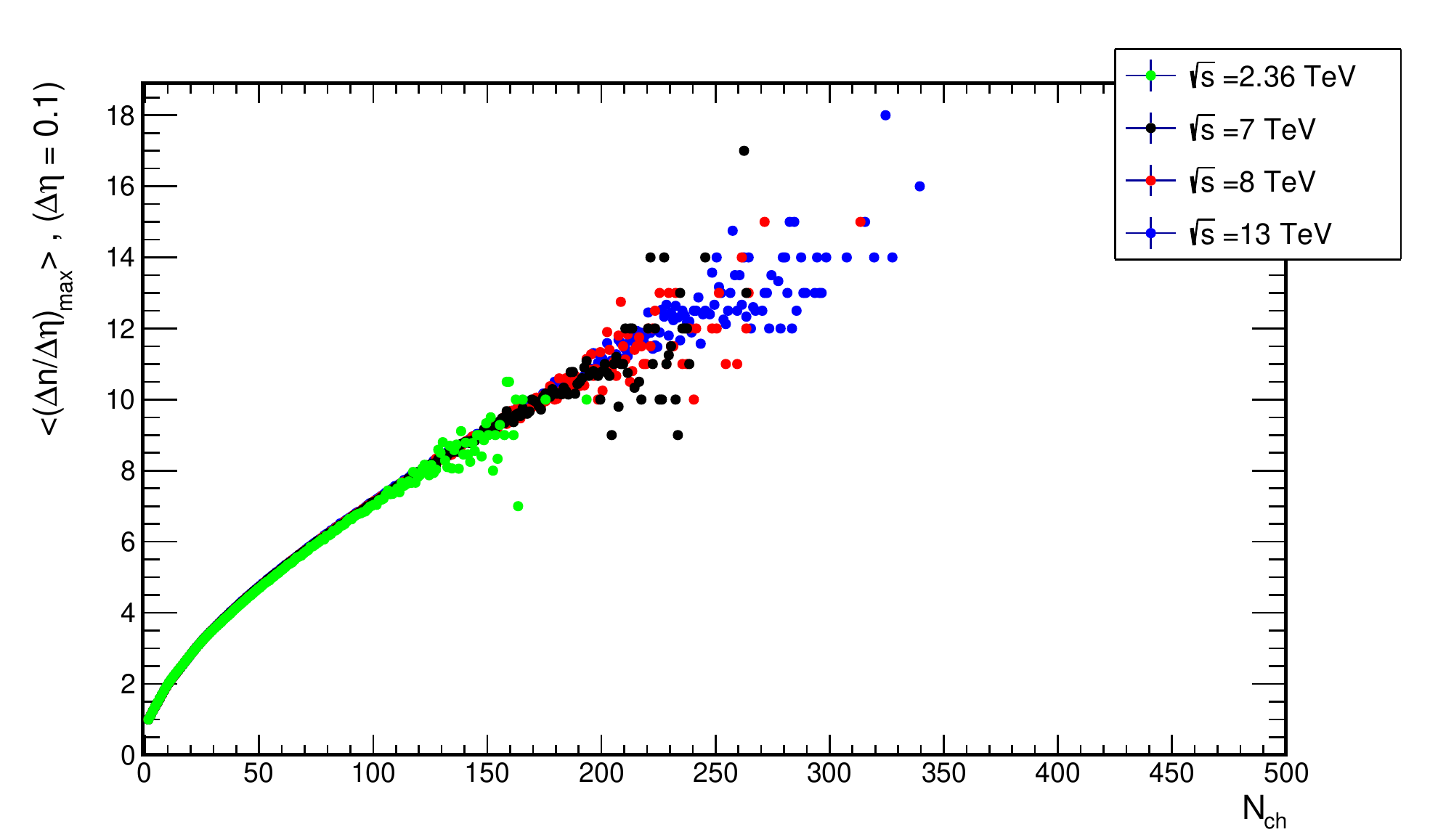}
\caption{ \footnotesize for HERWIG at $\Delta\eta = 0.1 $ } 
\end{subfigure}

%\begin{figure}[!tbp]
 \begin{subfigure}{0.49\textwidth}
    \includegraphics[width=\textwidth , height=5.8cm]{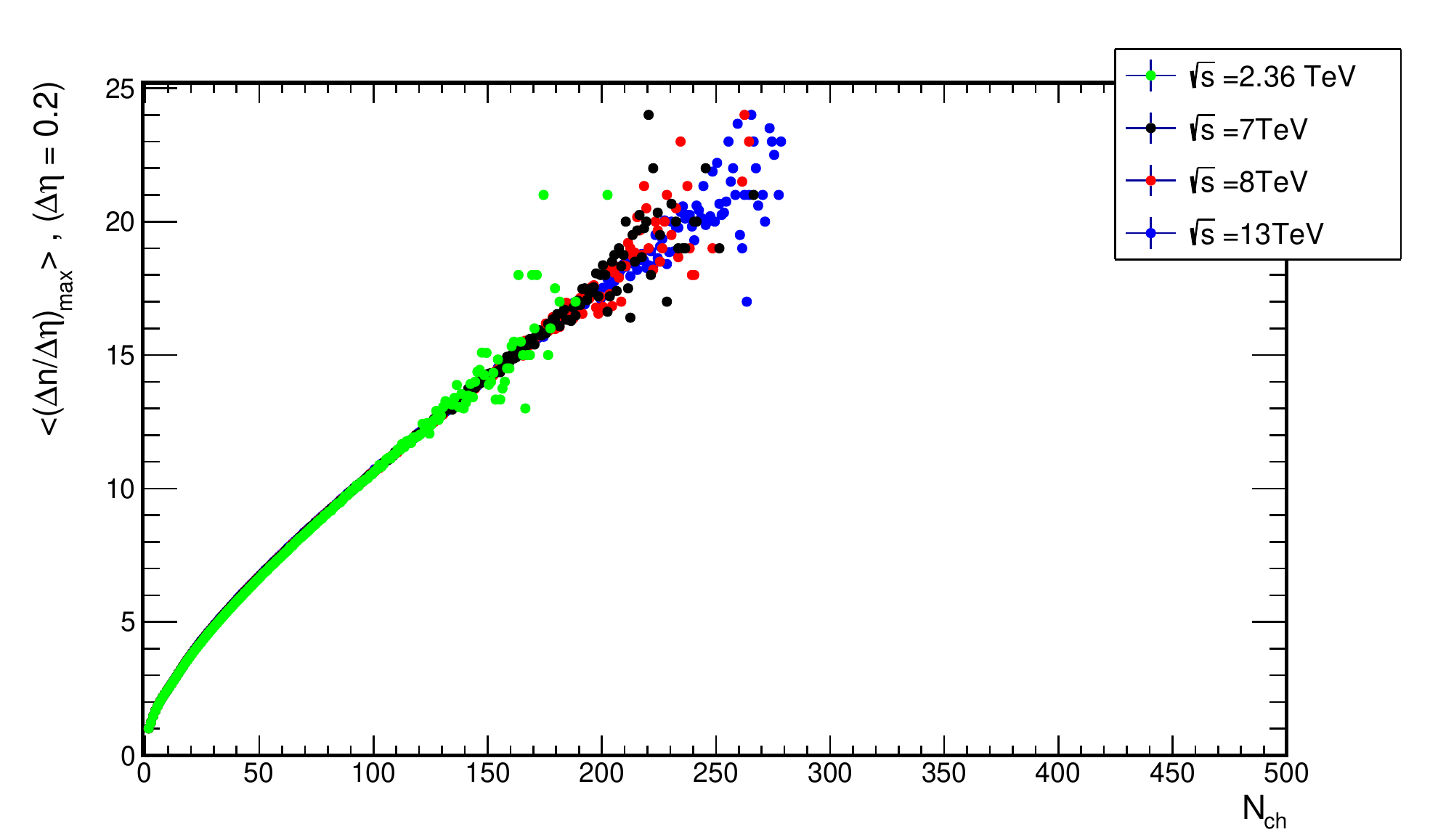}
\caption{\footnotesize for PYTHIA at $\Delta\eta = 0.2 $ } 
\end{subfigure}
  \hfill
   \begin{subfigure}{0.49\textwidth}
    \includegraphics[width=\textwidth , height=5.8cm]{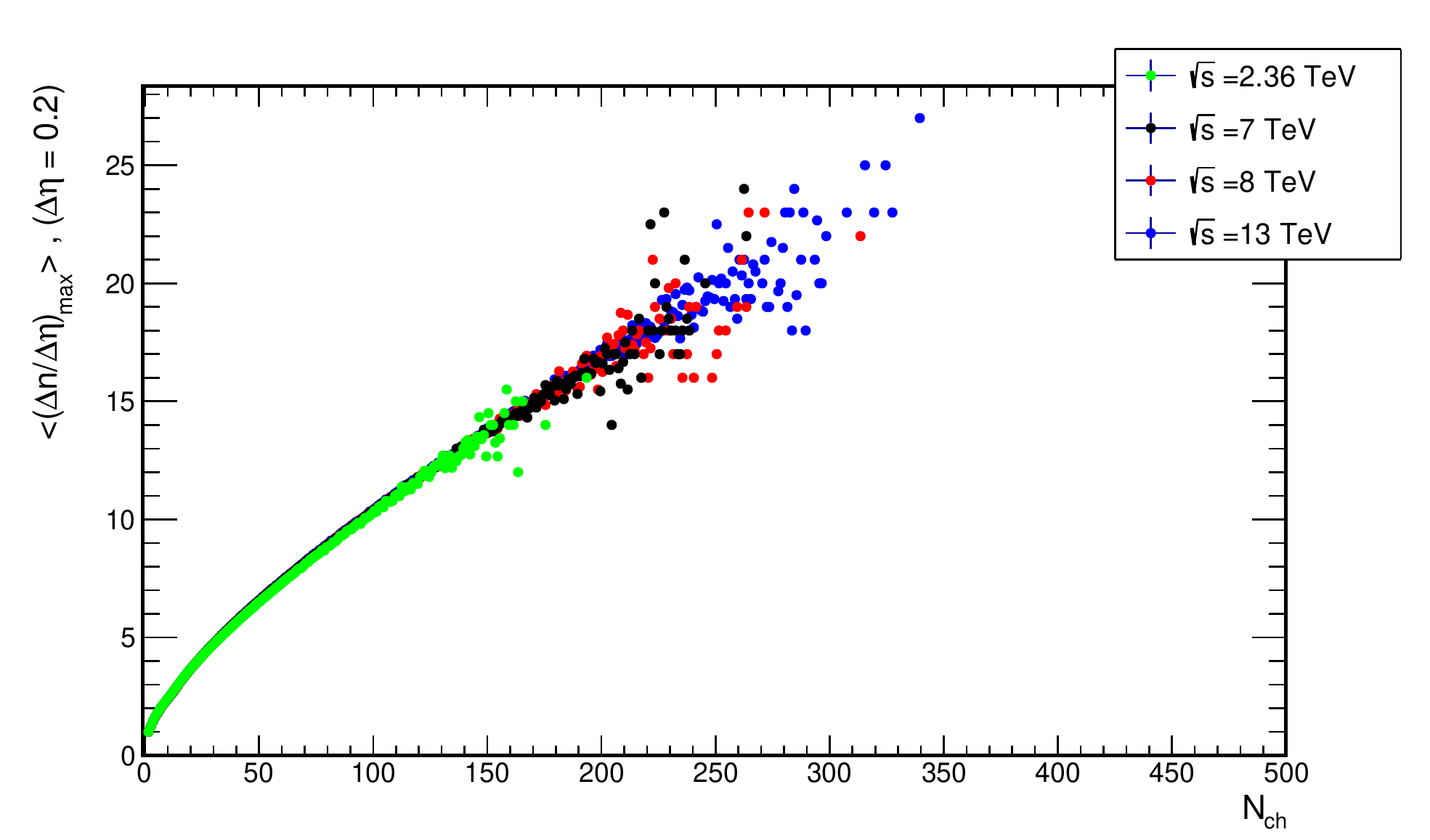}
\caption{\footnotesize for HERWIG at $\Delta\eta = 0.2 $ } 
\end{subfigure}

%\begin{figure}[!tbp]
 \begin{subfigure}{0.49\textwidth}
    \includegraphics[width=\textwidth , height=5.8cm]{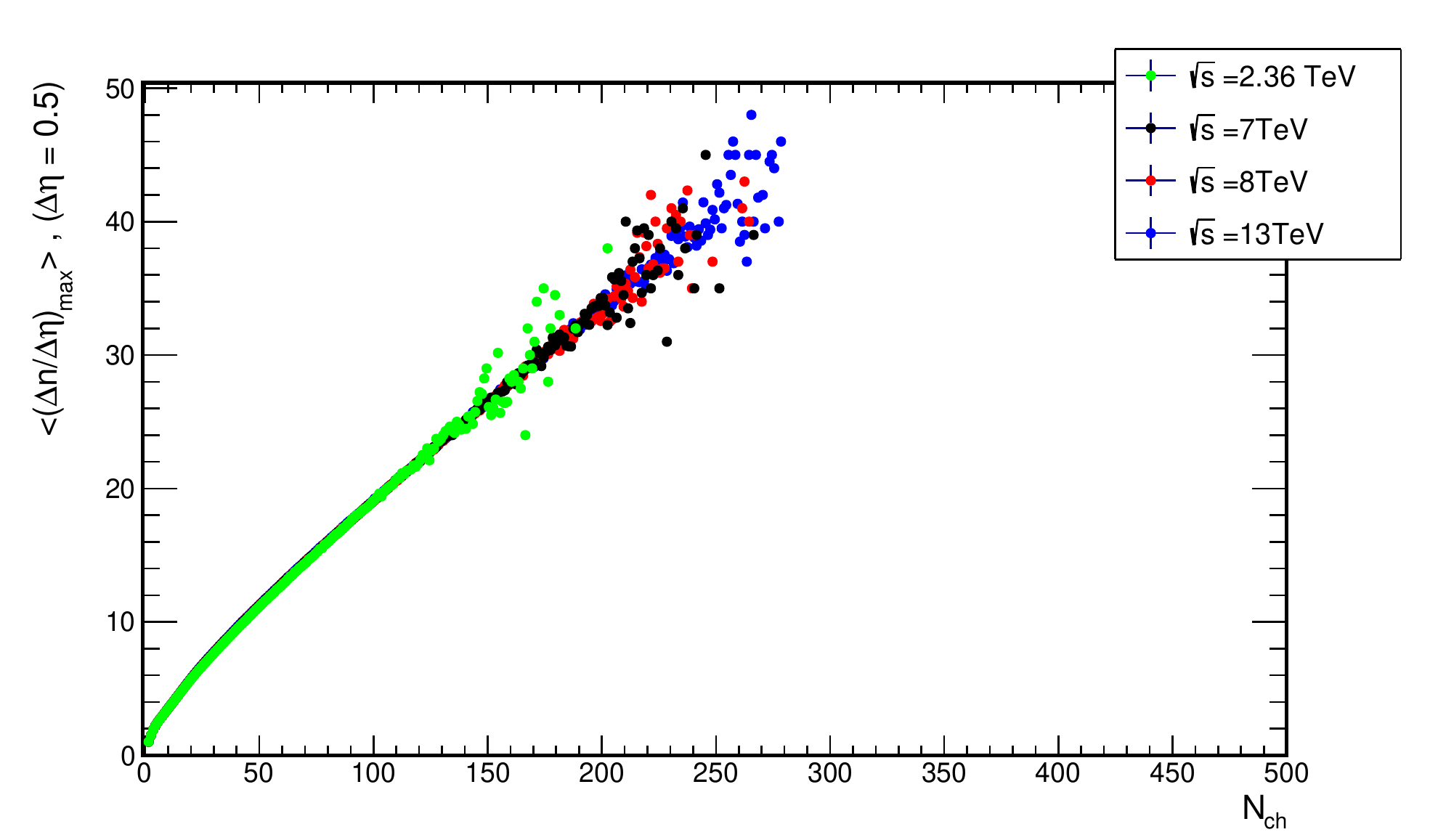}
\caption{\footnotesize for PYTHIA at $\Delta\eta = 0.5 $ } 
\end{subfigure}
  \hfill
   \begin{subfigure}{0.49\textwidth}
    \includegraphics[width=\textwidth , height=5.8cm]{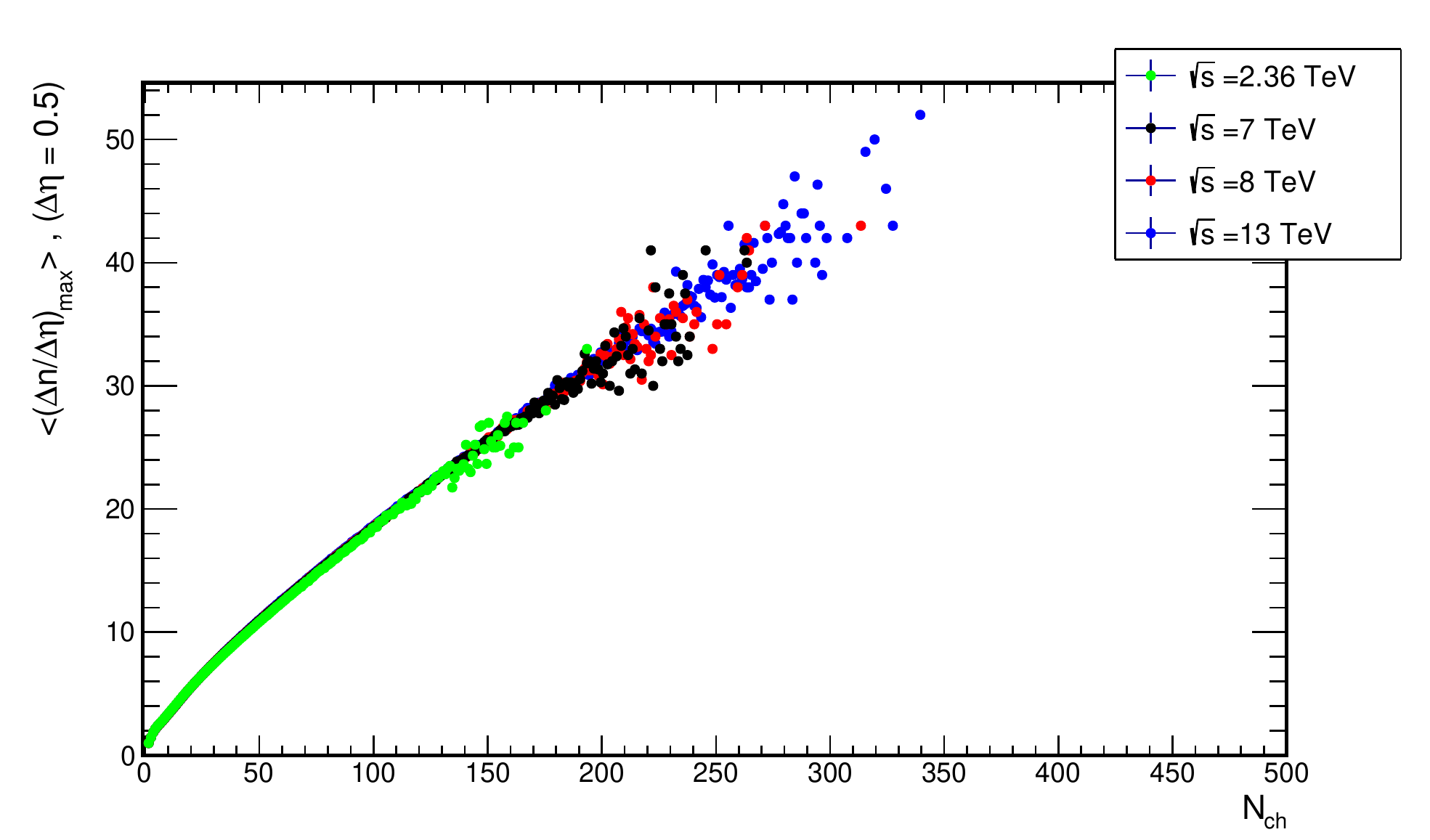}
\caption{\footnotesize for HERWIG at $\Delta\eta = 0.5 $ } 
\end{subfigure}
\caption{ (a), (c) and (e) represents $<(\Delta n / \Delta\eta)_{max}>$ as a function of $N_{ch}$ for $\Delta\eta = 0.1, 0.2 $ and $0.5$ respectively in $pp$ collisions at   $\sqrt{s} = 2.36$ TeV, $7$ TeV, $8$ TeV, $13$ TeV  using PYTHIA and (b), (d) and (f) are for HERWIG.} 
\end{figure}

\clearpage

\section{Conclusion}

In hadron-hadron collisions, events with a large number of charged particles within very narrow pseudorapidity windows have been observed by previous experiments.
Such analysis should  be conducted at the LHC in particular 
due to the availability of a large minimum bias data sets selected 
at the trigger level with high total 
charged-particle multiplicities.  These
trigger selection cuts will not significantly affect such a study.

In this paper, we show the baseline expectation, as predicted by two 
phenomenological models( PYTHIA and HERWIG), for such a typical analysis.
The models predict that 
the exponential behavior of the $dN/dn_{ch}$ will be expected for multi-particle production in 
the high energy regime of the LHC, and can be used as a 
monitor of the shape of the distribution to define a search region for anomalous events. 

However establishing any new 
effects based on one or just a few events, as often done in 
previous experiments, will not be an option.
But we expect the maximum track density method to be an effective technique in searching for anomalous events in the LHC data when complemented with an 
analysis in the $\phi$ dimension. Also, the average of the maximum number of charged particles has an almost linear dependence on the total multiplicity and approximately  independent on the energy up to multiplicities of 120, and will extend the previous work from a center of mass energy of 22 GeV up to 13-14 TeV.

\end{document}